\documentclass[aps,prx,superscriptaddress,nofootinbib,longbibliography]{revtex4-2}
\usepackage{amsmath,amssymb,amsfonts}
\usepackage{graphicx}
\usepackage{hyperref}
\usepackage{xcolor}

\usepackage{amsthm}

\usepackage{enumitem}

\usepackage{fancyhdr}
\usepackage[calcwidth]{titlesec}
\usepackage{setspace}

\usepackage{extarrows}

\usepackage[most]{tcolorbox}
\tcbuselibrary{breakable,skins}
\usepackage{xparse} % for optional arguments in \bmyt, \bmyl, \bmyd

\newtcolorbox{naturethmbox}[1][]{%
  enhanced,
  breakable,
  sharp corners,
  boxrule=0pt,
  frame hidden,
  colback=black!4.5,
  borderline west={0.7pt}{0pt}{black!50},
  left=7pt,
  right=7pt,
  top=5pt,
  bottom=5pt,
  before skip=8pt,
  after skip=8pt,
  #1
}

\newtcolorbox{naturedefbox}[1][]{%
  enhanced,
  breakable,
  sharp corners,
  boxrule=0pt,
  frame hidden,
  colback=black!4.5,
  borderline west={0.6pt}{0pt}{black!40},
  left=7pt,
  right=7pt,
  top=5pt,
  bottom=5pt,
  before skip=8pt,
  after skip=8pt,
  #1
}

\NewDocumentCommand{\bmyt}{o}{%
  \begin{naturethmbox}%
  \IfNoValueTF{#1}{\begin{theorem}}{\begin{theorem}[#1]}%
}
\newcommand{\emyt}{%
  \end{theorem}%
  \end{naturethmbox}%
}

\NewDocumentCommand{\bmyl}{o}{%
  \begin{naturethmbox}%
  \IfNoValueTF{#1}{\begin{lemma}}{\begin{lemma}[#1]}%
}
\newcommand{\emyl}{%
  \end{lemma}%
  \end{naturethmbox}%
}

\NewDocumentCommand{\bmyd}{o}{%
  \begin{naturedefbox}%
  \IfNoValueTF{#1}{\begin{definition}}{\begin{definition}[#1]}%
}
\newcommand{\emyd}{%
  \end{definition}%
  \end{naturedefbox}%
}

\NewDocumentCommand{\bmyc}{o}{%
  \begin{naturethmbox}%
  \IfNoValueTF{#1}{\begin{corollary}}{\begin{corollary}[#1]}%
}
\newcommand{\emyc}{%
  \end{corollary}%
  \end{naturethmbox}%
}

\usepackage{dsfont}

\usepackage{mathtools}

\def\1{\mathbf{1}}

\def\diag{{\rm Diag}}

\def\pure{{\rm Pure}}
\def\pos{{\rm Pos}}
\def\cp{{\rm CP}}
\def\cptp{{\rm CPTP}}

\def\L{{\mathbb L}}

\newcommand{\epm}{\end{pmatrix}}
\newcommand{\bpm}{\begin{pmatrix}}

\renewcommand{\log}{{\operatorname{log}}}

\newcommand{\ebm}{\end{bmatrix}}
\newcommand{\bbm}{\begin{bmatrix}}

\def\C{\mathbb{C}}
\def\R{\mathbb{R}}
\def\N{\mathbb{N}}

\def\>{\rangle}
\def\<{\langle}

\def\id{\mathsf{id}}

\def\mE{\mathcal{E}}

\def\mF{\mathcal{F}}
\def\mN{\mathcal{N}}

\def\mL{\mathcal{L}}

\def\mS{\mathcal{S}}

\def\mT{\mathcal{T}}
\def\mV{\mathcal{V}}
\def\mU{\mathcal{U}}

\def\mQ{\mathcal{Q}}

\newcommand{\supp}{\operatorname{supp}}

   \newcommand{\rl}{\rangle\langle}

\renewcommand{\ge}{\geqslant}
\renewcommand{\le}{\leqslant}

\newcommand{\ben}{\begin{enumerate}}
\newcommand{\een}{\end{enumerate}}

\theoremstyle{definition}
\newtheorem{theorem}{Theorem}
\newtheorem{lemma}[theorem]{Lemma}
\newtheorem{corollary}[theorem]{Corollary}

\newtheorem{definition}[theorem]{Definition}
\newtheorem{remark}[theorem]{Remark}

\usepackage{array}

\newcommand{\bea}{\begin{eqnarray}}
\newcommand{\eea}{\end{eqnarray}}
\newcommand{\be}{\begin{equation}}
\newcommand{\ee}{\end{equation}}
\newcommand{\ba}{\begin{equation}\begin{aligned}}
\newcommand{\ea}{\end{aligned}\end{equation}}

\newcommand{\bee}{\begin{enumerate}}
\newcommand{\eee}{\end{enumerate}}

\newcommand{\mA}{\mathcal{A}}

\newcommand{\ml}{\mathfrak{L}}
\newcommand{\mb}{\mathfrak{B}}
\newcommand{\md}{\mathfrak{D}}

\newcommand{\mR}{\mathcal{R}}
\newcommand{\mM}{\mathcal{M}}

\newcommand{\D}{\mathbf{D}}

\newcommand{\lr}{\rangle\langle}
\newcommand{\la}{\langle}
\newcommand{\ra}{\rangle}
\newcommand{\tr}{{\rm Tr}}

\newcommand{\eps}{\varepsilon}

\newcommand{\ket}[1]{|#1\rangle}

\newcommand{\eqdef}{\coloneqq}

\def\0{\mathbf{0}}

\def\tA{\tilde{A}}

\def\tD{\widetilde{D}}

\def\trho{{\tilde{\rho}}}

\def\d{{\mathrm{d}}}

\begin{document}

\title{Failure of the Asymptotic Equipartition Property for Quantum Channels}

\author{Gilad Gour}
\affiliation{Faculty of Mathematics, Technion--Israel Institute of Technology, Haifa 3200003, Israel}

\begin{abstract}
We establish two results on the asymptotic equipartition property
(AEP) for quantum channels. First, the AEP formulated using
smoothing over completely positive trace-preserving maps fails
in general. Second, the AEP with smoothing over subchannels
(completely positive trace-nonincreasing maps) is equivalent
to the strong converse for parallel quantum channel Stein's lemma,
to a sharp threshold for the channel hockey-stick divergence,
and to an AEP for the channel Lorenz divergence. The counterexample
reveals a surprising separation between channel and subchannel
smoothing. Although states and classical channels admit a sharp,
dimension-independent completion bound, which we establish,
the corresponding single-shot gap for quantum channels is unbounded
already for qubits. This separation persists asymptotically:
we construct qutrit channel pairs with finite max-relative entropy
whose CPTP-smoothed rate strictly exceeds the regularized channel
relative entropy. The asymptotic gap between channel and subchannel
smoothing is unbounded across the family, even when the subchannel
approximation error decays exponentially. To establish the
equivalences, we develop a uniform filter that converts Lorenz
smoothing into a single nearby subchannel. The four conjectured
formulations thus share a common asymptotic threshold, whose
validity for general channel pairs remains open. Together, these
results separate this threshold question from the obstruction
imposed by exact trace preservation.
\end{abstract}

\maketitle

\tableofcontents

\section{Introduction}
\label{sec:introduction}

The asymptotic equipartition property (AEP) connects one-shot
information measures to the entropy rates governing many independent
copies of a resource. In quantum information theory, smooth min-
and max-entropies converge to von Neumann entropies, and the smooth
max-relative entropy converges to the Umegaki relative
entropy~\cite{Renner2005,Datta2009,TCR2009,DR2009,DMHB2013,Tomamichel2016}.
Together with quantum Stein's lemma~\cite{HP1991,ON2000}, these
results connect approximation, distinguishability, and asymptotic
conversion. They are fundamental to quantum resource theories,
where smoothing represents an allowance for a small
error~\cite{CG2019,Gour2025}.

For states, allowing subnormalized smoothers does not change the
asymptotic rate~\cite{DR2009,Tomamichel2016,RLD2026}.
Normalization is a scalar constraint: if $\trho\le\gamma\sigma$
has trace $1-t$, adding $t\sigma$ restores its missing mass.
The resulting increase in max-relative entropy is bounded by a
dimension-independent constant at fixed smoothing error.
The same argument applies separately to each input of a classical
channel. It is therefore natural to expect channel and subchannel
smoothing to become equivalent in the many-copy limit.

We show that this expectation fails for quantum channels.
Our first result is a counterexample: an exponentially small
trace deficit can reduce the asymptotic smoothing rate, with
an unbounded gap across a family of qutrit channel pairs.
Consequently, the AEP with smoothing over completely positive
trace-preserving (CPTP) maps fails even when the unsmoothed
max-relative entropy is finite. Our second result identifies
what remains true at the level of formulations: the AEP with
smoothing over subchannels, namely completely positive
trace-nonincreasing maps, is equivalent to the strong converse
for parallel channel Stein's lemma, a sharp hockey-stick threshold,
and an AEP for the channel Lorenz divergence. The validity of
these four equivalent conjectures for general channel pairs
remains open.

To state the problem precisely, recall that channel divergences
optimize distinguishability over inputs, while their amortized
versions account for distinguishability already present in the
input states~\cite{WBHK2020}. The relative-entropy chain rule
identifies amortization with regularization~\cite{FFRS2020}:
\be\label{eq:intro-relative-entropy-rate}
D^{\mA}(\mN\|\mM)=D^{\rm reg}(\mN\|\mM)\,.
\ee
Writing $D_{\max}^{\delta}$ for the
max-relative entropy smoothed over CPTP maps within
diamond distance $\delta$, this gives the candidate asymptotic rate in the CPTP-smoothed AEP:
\be\label{eq:intro-CPTP-AEP}
\lim_{n\to\infty}\frac1nD_{\max}^{\delta}
\left(\mN^{\otimes n}\middle\|\mM^{\otimes n}\right)
\stackrel{?}{=}D^{\rm reg}(\mN\|\mM)\,,
\qquad\delta\in(0,1)\,.
\ee
In~\cite{WW2019}, this smoothing problem was related to approximate
distinguishability cost, leaving its asymptotic equality with
the regularized relative entropy open.
Positive AEP results for channel entropy and
max-information~\cite{GW2021,FWTB2020} concern more structured
settings. Strong Stein results hold for classical and
classical-quantum channels~\cite{Hayashi2009,WBHK2020}
and replacer alternatives~\cite{CMW2016}; restricted channel
AEPs also extend to von Neumann algebras~\cite{FGR2025}.
Channel chain rules and quantitative parallelization further
clarify the vanishing-error discrimination
rate~\cite{BT2022,BDSW2024}.

The connection to the strong converse requires care about what
is being smoothed. Connections between channel Stein properties,
output-smoothing AEPs, and continuity of regularized divergences
were established in~\cite{FGW2025}. Smoothing each output state
separately, however, need not produce a single linear map that
approximates the channel on every input. Requiring that map to
preserve trace introduces a further constraint. Our results
distinguish these three approximation problems and explain why
failure of~\eqref{eq:intro-CPTP-AEP} does not itself settle the
channel strong converse.

These distinctions also matter when drawing on the broader state
literature. The information-spectrum approach~\cite{NH2007}
and its smooth-entropy formulation~\cite{DR2009} treat general
state sequences. One-shot and second-order comparisons were
developed in~\cite{TH2013,DL2015}, while the relations between
testing, positive-part thresholds, and state smoothing were
sharpened in~\cite{RLD2026}. These tools underlie the threshold
approach used here, but do not directly supply a common channel
smoother for all entangled inputs.

The connection between testing and smoothing makes generalized
state Stein lemmas a natural starting point for the channel AEP.
The formulation in~\cite{BP2010} and its consequences for reversible
resource theories~\cite{BG2015} illustrate how asymptotic testing
can govern resource conversion. Following the identification of
a gap in the original proof~\cite{BBGLPRT2023}, independent
proofs were supplied in~\cite{HY2025,Lami2025}.
Extensions cover almost-i.i.d. null
hypotheses~\cite{Lami2025,MSR2026} and enable universal resource
distillation under asymptotically resource non-generating
operations~\cite{LRT2026}.
Related composite and adversarial testing results address
uncertainty in the hypotheses~\cite{BBH2021,MSW2022},
including settings with restricted measurements~\cite{BHLP2020}.
Particularly relevant here, the generalized AEP
of~\cite{FFF2026a} connects testing and smoothing for convex
state sets under suitable structural assumptions, including
tensor stability of the sets and their polars.
Under suitable semidefinite descriptions, the associated
regularized entropies can also be efficiently
approximated~\cite{FFF2026b}.
For channel smoothing, however, the approximations must arise
from one map acting consistently on every input. Our counterexample
shows why its trace-preserving constraint must also be accounted
for when seeking a channel AEP from these state-set results.

At the channel level, the connection depends further on who
chooses the inputs and which error limit is considered.
Composite channel discrimination~\cite{BDS2023} and generalized
Stein lemmas for classical--quantum dynamical
resources~\cite{HY2025b,BDK2025} extend testing beyond a fixed
pair of states. Adversarial models assign input choices
to an opponent~\cite{FFF2025,HCG2026}, while the minimax
framework of~\cite{FCCGH2026} gives separate inputs to a tester
and a jammer. Its general entangled-tester theorem takes the
limit of vanishing type-I error, while its fixed-error
strong-converse results concern more restricted settings.
These results do not establish the assertion relevant here:
the fixed-error strong converse for arbitrary channel pairs
with unrestricted parallel tester-chosen inputs.
Recent strong converses for communication
capacities~\cite{CT2026,BT2026} address a different operational
question, as do generalized quantum Sanov results with a
different orientation of composite
testing~\cite{LBR2026,Lami2025b} and related classical
results~\cite{Lami2025c}.
Our equivalence theorem connects the parallel discrimination
threshold to subchannel smoothing, while the counterexample
separates this threshold question from the obstruction imposed
by exact trace preservation.

Our starting point is the channel Lorenz curve, represented by
the hockey-stick divergences $E_\gamma(\mN\|\mM)$.
Relative Lorenz curves encode binary testing
tradeoffs~\cite{BG2017}, while integral representations connect
them to relative entropy and other quantum
divergences~\cite{Frenkel2023,HT2024,LHC2025}.
We denote by $D_{\max}^{\eps,\L}$ the quantum channel Lorenz
extension of the classical smoothed max-relative
entropy~\cite{Gour2026}, with values clipped at zero.
This notation emphasizes the order of construction:
first smoothing, then Lorenz extension.
Equivalently, it is the smallest $\log \gamma$,
with $\gamma\ge1$, for which $E_\gamma(\mN\|\mM)\le\eps$.
Writing $D_{\max}^{\delta,\le}$ for the max-relative entropy
smoothed over subchannels with error tolerance $\delta$, we prove
\be
D_{\max}^{\delta,\le}(\mN\|\mM)
\le D_{\max}^{\eps,\L}(\mN\|\mM)
\qquad\text{where}\qquad
\delta\eqdef4\sqrt{\eps}+2\eps<1\,.
\ee
This bound provides the link between the Lorenz description
and approximation by subchannels that underlies our AEP
equivalence theorem.

To obtain a channel approximation, the resulting subchannel
must be completed to a trace-preserving map without substantially
increasing its max-relative entropy. We establish a sharp,
dimension-independent bound on this increase for states and
classical channels, and extend it to structured quantum settings.
For general quantum channels, however, completion can incur an
unbounded cost already for qubits, even when the original channel
pair has finite max-relative entropy.

Our first main result shows that this obstruction persists
asymptotically.  We consider the qutrit channels
\be
\mN=\id_3\,,
\qquad
\mM=t\,\id_3+(1-t)\mU\,,
\qquad
U=\diag(1,e^{i\theta},e^{-i\theta})\,,
\ee
with parameters and smoothing radius as specified in
Section~\ref{subsec:asymptotic-qutrit-separation}, we prove
\ba\label{eq:intro-asymptotic-separation}
\lim_{n\to\infty}\frac1nD_{\max}^{\delta}
\left(\mN^{\otimes n}\middle\|\mM^{\otimes n}\right)
&=\log\frac1t\,,\\
\limsup_{n\to\infty}\frac1nD_{\max}^{\delta,\le}
\left(\mN^{\otimes n}\middle\|\mM^{\otimes n}\right)
&\le R_t\eqdef\frac12\log\frac1t+\log\frac4{1-t}\,.
\ea
The resulting gap is at least
\be
\log \frac1t-R_t=\log\frac{1-t}{4\sqrt t}\,,
\ee
which diverges as $t\downarrow0$ at fixed allowed $\theta$
and $\delta$. Each pair has finite $D_{\max}$, and the
constructed subchannels have exponentially vanishing generalized
diamond error. The direct part of state Stein's lemma implies
$D^{\rm reg}(\mN\|\mM)\le R_t<\log (1/t)$, disproving both
\eqref{eq:intro-CPTP-AEP} and its vanishing-error version.

Our second main result identifies four equivalent formulations
of the remaining AEP question. For pairs with finite $D_{\max}$,
Theorem~\ref{4conjectures} equates the subchannel-smoothed AEP
with the strong converse for parallel channel Stein's lemma,
the Lorenz-divergence AEP,
and the above-threshold assertion
\be
E_{2^{nr}}
\left(\mN^{\otimes n}\middle\|\mM^{\otimes n}\right)
\xrightarrow{n\to\infty}0
\qquad\forall \,r>D^{\mA}(\mN\|\mM)\,.
\ee
Below this threshold, convergence to one holds unconditionally.
The testing/output-smoothing connection is known~\cite{FGW2025};
our filter realizes the threshold through a single dominated
subchannel, and the Lorenz integral supplies the fourth
formulation. The strong converse allows arbitrary entangled
inputs and a reference system, without requiring exponential
convergence. The theorem proves equivalence, leaving the common
threshold question open for general channel pairs.

The paper follows this progression.
Section~\ref{sec:preliminaries} fixes notation.
Section~\ref{sec:channel-vs-subchannel-smoothing} develops
Lorenz filtering and positive completion results.
Section~\ref{sec:quantum-completion-gaps} gives the single-shot
and asymptotic counterexamples.
Section~\ref{sec:equivalent-AEP-formulations} proves the four
equivalences, and Section~\ref{sec:failure-CPTP-smoothed-AEP}
deduces failure of the CPTP-smoothed AEP.
Section~\ref{sec:discussion} discusses the remaining questions.

\section{Preliminaries}
\label{sec:preliminaries}

All systems are finite dimensional, logarithms are base two, and
$\tA\simeq A$. We write $\ml(A)$, $\pos(A)$, $\md(A)$, and
$\pure(A)$ for operators, positive operators, states, and pure
states, respectively, identifying pure states with their density
operators. We use $\inf\varnothing=+\infty$ and $-\log (0)=+\infty$.

\subsection{Channels, subchannels, and CP order}

The sets $\cp(A\to B)$, $\cp_\le(A\to B)$, and
$\cptp(A\to B)$ consist of completely positive maps,
completely positive trace-nonincreasing maps, and completely
positive trace-preserving maps, respectively. We call elements of
$\cp_\le(A\to B)$ subchannels. If $\mE^*$ is the adjoint with
respect to the Hilbert--Schmidt inner product, then
\be
\mE\in\cp_\le(A\to B)
\quad\Longleftrightarrow\quad
\mE\in\cp(A\to B)\,,\quad\mE^*(I_B)\le I_A\,,
\ee
with equality in the last condition for channels. For two
completely positive maps, $\mE\le\mF$ means that $\mF-\mE$
is completely positive. We refer to this as CP order.

We use the unnormalized Choi operator~\cite{Choi1975,Watrous2018}
\be\label{eq:prelim-Choi}
J_{\!AB}^{\mE}
\eqdef\sum_{i,j}|i\rl j|_A\otimes
\mE_{\tA\to B}(|i\rl j|_{\tA})\,,
\ee
where the bases of $A$ and $\tA$ are identified. Thus CP order is
equivalent to the corresponding order of Choi operators, and a
completely positive map is trace preserving precisely when
$\tr_B[J_{\!AB}^{\mE}]=I_A$.
For a Hermiticity-preserving map $\mE$, the diamond norm can be
written as~\cite{Watrous2018}
\be\label{eq:prelim-diamond}
\|\mE\|_\diamond
=\max_{\psi\in\pure(RA)}
\big\|(\id_R\otimes\mE)(\psi)\big\|_1\,,
\qquad R\simeq A\,.
\ee
For $\mE\in\cp(A\to B)$ it satisfies
$\|\mE\|_\diamond=\|\mE^*(I_B)\|_\infty$.
Channel smoothing uses $\tfrac12\|\mN-\mM\|_\diamond$; the
subchannel distance is defined in
Section~\ref{sec:channel-vs-subchannel-smoothing}.

\subsection{State and channel divergences}

We use the Umegaki relative entropy
$D(\rho\|\sigma)=\tr[\rho(\log \rho-\log \sigma)]$, with value
$+\infty$ unless $\supp(\rho)\subseteq\supp(\sigma)$.
The max-relative entropy is~\cite{Datta2009}
\be\label{eq:prelim-state-Dmax}
D_{\max}(\rho\|\sigma)
\eqdef\inf\{\log \gamma:\gamma>0,\ \rho\le\gamma\sigma\}\,.
\ee
The latter definition also applies to positive semidefinite,
possibly subnormalized operators.

A quantum divergence $\D$ obeys the data-processing inequality
under channels. Its channel extension is obtained by optimizing
over inputs together with a reference
system~\cite{WBHK2020,FGW2025}:
\be\label{eq:prelim-channel-extension}
\D(\mN\|\mM)
\eqdef\sup_{\psi\in\pure(A\tA)}
\D\big(\mN_{\tA\to B}(\psi_{A\tA})
\big\|\mM_{\tA\to B}(\psi_{A\tA})\big)\,.
\ee
The identity on the reference is implicit. Purification and data
processing allow the restriction to $\pure(A\tA)$.

For max-relative entropy, the extension has the CP-order form
\be\label{eq:prelim-channel-Dmax}
D_{\max}(\mN\|\mM)
=\inf\{\log \gamma:\gamma>0,\ \mN\le\gamma\mM\}\,.
\ee
We use the same CP-order definition when the first map is a
subchannel. In finite dimensions, $D_{\max}(\mN\|\mM)<\infty$
is equivalent to $\supp(J^{\mN})\subseteq\supp(J^{\mM})$.
For channels, max-relative entropy is additive under tensor
products~\cite{WW2019}.

The amortized divergence associated with $\D$ is~\cite{WBHK2020}
\be\label{eq:prelim-amortized}
\D^{\mA}(\mN\|\mM)
\eqdef\sup_{\substack{R\\\rho,\sigma\in\md(RA)}}
\Big\{\D\big(\mN_{A\to B}(\rho_{RA})
\big\|\mM_{A\to B}(\sigma_{RA})\big)
-\D(\rho_{RA}\|\sigma_{RA})\Big\}\,,
\ee
where pairs with infinite input divergence are excluded. Its
regularized channel counterpart, whenever the limit exists, is
\be\label{eq:prelim-regularized}
\D^{\rm reg}(\mN\|\mM)
\eqdef\lim_{n\to\infty}\frac1n
\D\left(\mN^{\otimes n}\middle\|\mM^{\otimes n}\right)\,.
\ee
For $\D=D$, superadditivity and Fekete's lemma give existence of
the limit and identify it with the supremum of the normalized
block divergences~\cite{FFRS2020}. If $D_{\max}(\mN\|\mM)<\infty$, this limit
is finite. The relative-entropy chain rule gives~\cite{FFRS2020}
\be\label{eq:prelim-amortization-equality}
D^{\rm reg}(\mN\|\mM)=D^{\mA}(\mN\|\mM)\,.
\ee

\subsection{Hypothesis testing and the strong converse}
\label{subsec:prelim-hypothesis-testing}

For $\eps\in[0,1)$, set~\cite{TH2013,DMHB2013}
\be
\beta_{\rho|\sigma}(\eps)
\eqdef\min_{\substack{0\le\Lambda\le I\\
\tr[\Lambda\rho]\ge1-\eps}}\tr[\Lambda\sigma]\qquad\text{and}\qquad
D_{\rm H}^{\eps}(\rho\|\sigma)
\eqdef-\log\, \beta_{\rho|\sigma}(\eps)\,.
\ee
We set $D_{\rm H}^{1}=+\infty$.
The channel quantity $D_{\rm H}^{\eps}(\mN\|\mM)$ is its
extension~\eqref{eq:prelim-channel-extension}.

We consider parallel strategies with arbitrary entangled inputs
$\psi_{RA^n}$ and joint tests $0\le\Lambda_n\le I_{RB^n}$,
whose errors are~\cite{WBHK2020,FGW2025}
\be\label{eq:prelim-channel-errors}
\alpha_n\eqdef1-\tr\!\left[\Lambda_n
(\id_R\otimes\mN^{\otimes n})(\psi_{RA^n})\right]\qquad\text{and}\qquad
\beta_n\eqdef\tr\!\left[\Lambda_n
(\id_R\otimes\mM^{\otimes n})(\psi_{RA^n})\right]\,.
\ee
For states, quantum Stein's lemma states that~\cite{HP1991,ON2000}
\be\label{eq:prelim-state-Stein}
\lim_{n\to\infty}\frac1nD_{\rm H}^{\eps}
\left(\rho^{\otimes n}\middle\|\sigma^{\otimes n}\right)
=D(\rho\|\sigma)
\qquad\forall\,\eps\in(0,1)\,.
\ee
Applying its direct part to repeated blocks of channel inputs
gives the unconditional bound
\be\label{eq:prelim-channel-direct}
\liminf_{n\to\infty}\frac1nD_{\rm H}^{\eps}
\left(\mN^{\otimes n}\middle\|\mM^{\otimes n}\right)
\ge D^{\rm reg}(\mN\|\mM)\,;
\ee
we give the block argument in
Lemma~\ref{lem:aep-below-threshold}.
The corresponding strong-converse property would assert that every
sequence of parallel strategies satisfies
\be\label{eq:prelim-strong-converse}
\liminf_{n\to\infty}-\frac1n\log \beta_n
>D^{\rm reg}(\mN\|\mM)
\quad\Longrightarrow\quad
\alpha_n\longrightarrow1\,.
\ee
Unlike the direct bound~\eqref{eq:prelim-channel-direct}, this
property is not established here for general channel pairs.
Together with the direct bound, it is equivalent to the fixed-error
channel Stein limit appearing as statement~(1) of
Theorem~\ref{4conjectures}. That theorem proves the equivalence
of this statement with the subchannel-smoothed AEP and the other
formulations listed there; it does not establish that these
equivalent properties hold in general. Throughout, the strong
converse requires only $\alpha_n\to1$, without an exponential
convergence requirement.

\subsection{Hockey-stick divergences and Lorenz curves}

For normalized states and $\gamma\ge0$, define
\be\label{eq:prelim-state-hockey-stick}
E_\gamma(\rho\|\sigma)
\eqdef\tr(\rho-\gamma\sigma)_+
=\max_{0\le\Lambda\le I}
\big\{\tr[\Lambda\rho]-\gamma\tr[\Lambda\sigma]\big\}\,,
\ee
where $X_+$ is the positive part of a Hermitian operator $X$.
This is the support function of the binary testing region in
the direction $(1,-\gamma)$~\cite{BG2017}. Its channel extension is
\be\label{eq:prelim-channel-hockey-stick}
E_\gamma(\mN\|\mM)
\eqdef\max_{\psi\in\pure(A\tA)}
E_\gamma\big(\mN_{\tA\to B}(\psi_{A\tA})
\big\|\mM_{\tA\to B}(\psi_{A\tA})\big)\,.
\ee
It is nonincreasing in $\gamma$, lies in $[0,1]$, and satisfies
\be\label{eq:prelim-hockey-stick-order}
E_\gamma(\mN\|\mM)=0
\quad\Longleftrightarrow\quad\mN\le\gamma\mM\,.
\ee
In particular, for $\gamma\ge1$,
\be
E_\gamma(\mN\|\mM)
=\frac12\big(\|\mN-\gamma\mM\|_\diamond+1-\gamma\big)\,.
\ee
Convexity extends the state testing
dualities~\cite{BG2017,RLD2026} to give
\be\label{dual1}
E_\gamma(\mN\|\mM)
=\sup_{\eps\in[0,1]}
\left\{1-\eps-\gamma\,2^{-D_{\rm H}^{\eps}(\mN\|\mM)}\right\}
\ee
and, for $\eps\in(0,1)$,
\be\label{dual2}
2^{-D_{\rm H}^{\eps}(\mN\|\mM)}
=\sup_{\gamma>0}\frac{1-\eps-E_\gamma(\mN\|\mM)}{\gamma}\,.
\ee
Whenever
$E_\gamma(\mN\|\mM)<1-\eps$,
\be\label{eq:prelim-testing-bound}
D_{\rm H}^{\eps}(\mN\|\mM)
\le\log \gamma-\log \big(1-\eps-E_\gamma(\mN\|\mM)\big)\,.
\ee

Frenkel's integral representation of state relative entropy,
written in hockey-stick form, is~\cite{Frenkel2023,HT2024}
\be\label{eq:prelim-state-Frenkel}
D(\rho\|\sigma)
=\int_1^\infty\d\gamma
\left(\frac1\gamma E_\gamma(\rho\|\sigma)
+\frac1{\gamma^2}E_\gamma(\sigma\|\rho)\right)\,.
\ee
Replacing the state hockey-stick divergences by their channel
extensions defines the channel Lorenz divergence
\be\label{eq:prelim-channel-Lorenz}
D^\L(\mN\|\mM)
\eqdef\int_1^\infty\d\gamma
\left(\frac1\gamma E_\gamma(\mN\|\mM)
+\frac1{\gamma^2}E_\gamma(\mM\|\mN)\right).
\ee
Optimizing the input separately at each point of the integral gives
$D^\L(\mN\|\mM)\ge D(\mN\|\mM)$. No additivity or existence
of a regularized limit for $D^\L$ is assumed.

\section{Channel versus subchannel smoothing}
\label{sec:channel-vs-subchannel-smoothing}

In this section, we compare two natural smoothing conventions for the
channel max-relative entropy: smoothing over trace-preserving channels
and smoothing over trace-nonincreasing completely positive maps. We
first relate subchannel smoothing to the Lorenz-smoothed max-relative
entropy. We then show that, for states and classical channels, allowing
subnormalized smoothers provides at most a dimension-independent
additive advantage and therefore has no effect on asymptotic rates. The
same conclusion extends to several structured classes of quantum
channels. These positive results isolate the completion property that
fails for general quantum channels and prepare the counterexamples of
the following section.

For a Hermitian
operator $X$, define the generalized trace norm, following the
state smoothing convention in~\cite{Tomamichel2016,RLD2026},
\be
 \|X\|_+
 \eqdef
 \frac12\bigl(\|X\|_1+|\tr X|\bigr).
\ee
Its channel counterpart is the generalized diamond norm
\be\label{eq:generalized-diamond-norm}
 \|\mE\|_{\diamond,+}
 \eqdef
 \sup_{\psi\in\pure(RA)}
 \bigl\|(\id_R\otimes\mE)(\psi)\bigr\|_+,
 \qquad R\simeq A,
\ee
defined for every Hermiticity-preserving map $\mE:\ml(A)\to \ml(B)$.  Notice that
if $\mN$ and $\mM$ are channels, then
\be\label{eq:generalized-diamond-on-channels}
 \|\mN-\mM\|_{\diamond,+}
 =\frac12\|\mN-\mM\|_\diamond,
\ee
because the output difference has trace zero.  Thus the generalized
distance agrees with the usual channel distance on normalized maps, while
also accounting for trace loss when one of the maps is a subchannel.

For $\mN\in\cptp(A\to B)$ and $\delta\in(0,1)$, set
\be
 \mb^\delta(\mN)
 \eqdef
 \left\{\mN'\in\cptp(A\to B):
 \|\mN-\mN'\|_{\diamond,+}\le\delta\right\}
\ee
and
\be
 \mb_\le^\delta(\mN)
 \eqdef
 \left\{\mE\in\cp_\le(A\to B):
 \|\mN-\mE\|_{\diamond,+}\le\delta\right\}.
\ee
We use the notation
\be
 D_{\max}^{\delta}(\mN\|\mM)
 \eqdef
 \inf_{\mN'\in\mb^\delta(\mN)}D_{\max}(\mN'\|\mM)
\ee
and
\be
 D_{\max}^{\delta,\le}(\mN\|\mM)
 \eqdef
 \inf_{\mE\in\mb_\le^\delta(\mN)}D_{\max}(\mE\|\mM).
\ee
In view of~\eqref{eq:generalized-diamond-on-channels}, the first definition
is precisely the usual CPTP-smoothed max-relative entropy~\cite{GFW+2018,LW2019,FWTB2020,WW2019}.

\subsection{From Lorenz smoothing to subchannel smoothing}
\label{subsec:Lorenz-to-subchannel}

We next introduce a smoothing convention defined directly in terms of the
channel Lorenz curve. Recall that the channel hockey-stick divergence is
given by
\be
E_\gamma(\mN\|\mM)
\eqdef
\max_{\psi\in\pure(RA)}
E_\gamma\left(
(\id_R\otimes\mN)(\psi)
\middle\|
(\id_R\otimes\mM)(\psi)
\right),
\qquad R\simeq A,
\ee
where $\gamma\ge1$. Equivalently,
$E_\gamma(\mN\|\mM)$ measures the maximal violation, over all inputs
possibly entangled with a reference system, of the CP-order relation
$\mN\le\gamma\mM$. In particular,
\be
E_\gamma(\mN\|\mM)=0
\quad\Longleftrightarrow\quad
\mN\le\gamma\mM.
\ee

We use the channel Lorenz extension of the modified smooth
max-relative entropy $\widetilde D_{\max}^{\eps}$
of~\cite{RLD2026}, originally introduced as the upper
information-spectrum relative entropy $\overline D_s^\eps$
in~\cite[Definition~4.1]{DL2015}.
At the state level, this quantity is defined by thresholding
$\tr(\rho-\gamma\sigma)_+$; replacing the state hockey-stick
divergence by its channel counterpart gives the definition below.
Our restriction $\gamma\ge1$ clips the resulting quantity at zero.
We write $D_{\max}^{\eps,\L}$ to emphasize the order of
construction: We write $D_{\max}^{\eps,\L}$ to emphasize the order of construction: first smoothing at the classical state level then applying the unique
Lorenz extension to quantum channels~\cite{Gour2026}.

\bmyd[Lorenz-smoothed max-relative entropy]
\label{def:Lorenz-smoothed-max}
Let $\mN,\mM\in\cptp(A\to B)$ and let $\eps\in[0,1)$. We define
\be\label{eq:Lorenz-smoothed-max}
D_{\max}^{\eps,\L}(\mN\|\mM)
\eqdef
\inf\big\{
\log (\gamma)\;:\;
\gamma\ge1\,,\quad
E_\gamma(\mN\|\mM)\le\eps
\big\}.
\ee
\emyd

Unlike the smoothing quantities
$D_{\max}^{\delta}$ and $D_{\max}^{\delta,\le}$, the optimization
in~\eqref{eq:Lorenz-smoothed-max} is not over a neighborhood of
$\mN$. Instead, smoothing is performed directly on the Lorenz curve by
allowing a violation of size at most $\eps$ in the comparison
$\mN\le\gamma\mM$. At zero smoothing, the usual max-relative entropy
is recovered:
\be
D_{\max}^{0,\L}(\mN\|\mM)
=
D_{\max}(\mN\|\mM).
\ee
Moreover, $\eps\mapsto D_{\max}^{\eps,\L}(\mN\|\mM)$ is
nonincreasing.

The following theorem shows that Lorenz smoothing always produces a
nearby subchannel with no larger max-relative entropy. This implication
is universal: it requires no structural assumption on either channel.
Its state analogue is the approximate-domination smoothing lemma
of Datta and Renner~\cite{DR2009}, sharpened in~\cite{RLD2026}.
The channel proof below uses CP-order Radon--Nikodym theory to obtain
one linear approximation for all inputs~\cite{Raginsky2003}.

\bmyt
\label{thm:subchannel-vs-Lorenz-max}
Let $\mN,\mM\in\cptp(A\to B)$ and choose $\eps>0$ such that $\delta=4\sqrt{\eps}+2\eps<1$. Then
\be\label{eq:subchannel-vs-Lorenz-max}
D_{\max}^{\delta,\le}(\mN\|\mM)
\le
D_{\max}^{\eps,\L}(\mN\|\mM)\,.
\ee
\emyt

\begin{proof}
If the right-hand side of~\eqref{eq:subchannel-vs-Lorenz-max} is infinite,
the claim is immediate. Otherwise, let $\gamma=2^{D_{\max}^{\eps,\L}(\mN\|\mM)}$. Then, $\gamma\ge 1$  satisfies
$
s\eqdef E_\gamma(\mN\|\mM)\le\eps
$.
We show that there exists
$\mN_\gamma\in\cp_\le(A\to B)$ such that
$
\mN_\gamma\le\gamma\mM
$
and
\be\label{eq:subchannel-filter-distance}
\frac12\|\mN_\gamma-\mN\|_\diamond
\le
2\sqrt{s}+s\le\delta\,.
\ee
Semidefinite-program duality applied to the tester representation
 gives the following positive-part
representation (see~\cite{Watrous2018} for CP-map norm programs and
strong duality):
\be
E_\gamma(\mN\|\mM)
=
\min_{\substack{\mQ\in\cp(A\to B)\\
                \mQ\ge\mN-\gamma\mM}}
\|\mQ\|_\diamond\,,
\ee
whose dual is
\be
\begin{aligned}
\text{maximize}\quad
&\tr\left[J_{\!AB}^{\mN-\gamma\mM} \Gamma_{\!AB}\right]\\
\text{subject to}\quad
&0\le \Gamma_{\!AB}\le \rho_A\otimes I_B\\
&\rho\in\md(A)\,.
\end{aligned}
\ee
Choose an optimal completely positive map $\mQ$ such that
$
\|\mQ\|_\diamond=s
$
and
$
\mN\le\mS$,
where
$\mS\eqdef\gamma\mM+\mQ$.
Let $V_{\mM},V_{\mQ}:A\to BE$ be Stinespring operators~\cite{Stinespring1955} for
$\mM$ and $\mQ$, respectively. Then
\be
V_{\mS}
=
\sqrt{\gamma}\,V_{\mM}\otimes|0\ra^{E'}+ V_{\mQ}\otimes|1\ra^{E'}
\ee
is a Stinespring operator for $\mS$, where $E'$ is a two-dimensional system.
Since $\mN\le\mS$, the Radon--Nikodym theorem for completely positive
maps~\cite{Raginsky2003} implies that there exists a positive contraction (an effect)
$
0\le \Lambda\le I_{EE'}
$
such that
\be
W
\eqdef
\left(I_B\otimes \Lambda^{1/2}_{EE'}\right)V_{\mS}=X+Y
\ee
where
\be
X\eqdef\sqrt{\gamma}\left(I_B\otimes \Lambda^{1/2}_{EE'}\right)\left(V_{\mM}\otimes|0\ra^{E'}\right)\qquad\text{and}\qquad Y\eqdef \left(I_B\otimes \Lambda^{1/2}_{EE'}\right)\left(V_{\mQ}\otimes|1\ra^{E'}\right)
\ee
is a Stinespring operator for $\mN$. Since $\mN$ is trace preserving,
$
W^*W=I_A
$ and in particular
$\|W\|_\infty=1$.
The
contractivity of $\Lambda^{1/2}_{EE'}$ gives
\be\label{ybound}
\|Y\|_\infty\le\|V_{\mQ}\|_\infty=\sqrt{\|\mQ\|_\diamond}=\sqrt{s}\,.
\ee
Let $\widehat{\mN}_\gamma$ be the completely positive map with
Stinespring operator $X$. Since $X$ is obtained from the
$\sqrt{\gamma}V_{\mM}$ component by an environmental contraction, we have
$
\widehat{\mN}_\gamma\le\gamma\mM
$.

For an arbitrary ancilla $R$ and pure state $\psi\in\pure(RA)$, set
\be
\ket{w}\eqdef(I_R\otimes W)\ket{\psi},
\qquad
\ket{x}\eqdef(I_R\otimes X)\ket{\psi},\qquad|y\ra\eqdef\ket{w}-\ket{x}=(I_R\otimes Y)\ket{\psi}\,.
\ee
By contractivity of the trace norm on Hermitian operators under
the partial trace,
\ba\label{lis}
\left\|
(\id_R\otimes\mN)(\psi)
-
(\id_R\otimes\widehat{\mN}_\gamma)(\psi)
\right\|_1
&\le
\big\|
|w\lr w|-|x\lr x|
\big\|_1\\
&=\big\|
|w\lr y|+|y\lr x|
\big\|_1\\
&\le \big\|
|w\lr y|\big\|_1+\big\||y\lr x|
\big\|_1\\
&=\big\||y\ra\big\|\left(1+\big\||x\ra\big\|\right)\;,
\ea
where in the last equality we used the fact that $|w\ra$ is normalized.
Using  \eqref{ybound} we have $\||y\ra\|\le\sqrt{s}$ 
while the triangle inequality gives
\be
\big\||x\ra\big\|
\le
\big\||w\ra\big\|+\big\||y\ra\big\|
\le
1+\sqrt{s}\,.
\ee
Substituting these bounds into~\eqref{lis} and taking the supremum over all $\psi\in\pure(RA)$ yields
\be\label{scdist}
\|\mN-\widehat{\mN}_\gamma\|_\diamond
\le
2\sqrt{s}+s.
\ee

The map $\widehat{\mN}_\gamma$ need not be trace nonincreasing.
We therefore define
\be
\mN_\gamma
\eqdef
\frac1a\widehat{\mN}_\gamma
\qquad\text{where}\qquad
a
\eqdef
\max\left\{
1,\,
\|\widehat{\mN}_\gamma\|_\diamond
\right\}\,.
\ee
Since $\widehat{\mN}_\gamma$ is completely positive,
$\mN_\gamma$ is completely positive and trace nonincreasing, and
\be
\mN_\gamma
\le
\widehat{\mN}_\gamma
\le
\gamma\mM\,.
\ee
It is therefore left to estimate the trace-distance between $\mN$ and $\mN_\gamma$. From the triangle inequality 
\be\label{tria}
\|\mN_\gamma-\mN\|_\diamond\le\|\mN_\gamma-\widehat{\mN}_\gamma\|_\diamond+\|\widehat{\mN}_\gamma-\mN\|_\diamond
\ee
Moreover, observe that
\ba
\|\mN_\gamma-\widehat{\mN}_\gamma\|_\diamond&=a-1\\
&\le \|\widehat{\mN}_\gamma-\mN\|_\diamond\,,
\ea
since $\|\mN\|_\diamond=1$.
Combining this with~\eqref{tria} and then using~\eqref{scdist} yields
\be
\frac12\|\mN-\mN_\gamma\|_\diamond
\le \|\widehat{\mN}_\gamma-\mN\|_\diamond\le
2\sqrt{s}+s
\le
\delta\,.
\ee
Since $\|\mL\|_{\diamond,+}\le\|\mL\|_\diamond$ for a
Hermiticity-preserving map $\mL$, the preceding estimate gives
\be
\|\mN-\mN_\gamma\|_{\diamond,+}
\le4\sqrt{s}+2s\le\delta.
\ee
Hence
$
\mN_\gamma
\in
\mb_\le^{\delta}(\mN)$.
Since $\mN_\gamma\le\gamma\mM$, we also have
\be
D_{\max}(\mN_\gamma\|\mM)
\le
\log(\gamma).
\ee
Consequently,
\ba
D_{\max}^{\delta,\le}(\mN\|\mM)
&\le
\log(\gamma)\\
&=D_{\max}^{\eps,\L}(\mN\|\mM)\,.
\ea
This completes the proof of~\eqref{eq:subchannel-vs-Lorenz-max}.
\end{proof}

\subsection{When subnormalization is harmless}
\label{subsec:subnormalization-harmless}

For states and classical channels, the trace lost through smoothing
can be restored at a dimension-independent additive cost. The same
bound holds for several structured quantum settings. We first isolate
the completion property behind this behavior, before showing in the
next section how it fails for general quantum channels.

For states, a stronger relation is already known: with generalized
trace distance, the normalized smoothed max-relative entropy is the
maximum of zero and its subnormalized counterpart~\cite[Section~IV]{RLD2026}.
The following argument extends the elementary operation of restoring
missing probability mass to a sufficient completion condition for channels.
Given a subchannel $\mE$, its trace deficit is the effect
\be\label{eq:trace-deficit}
 \Delta_{\mE}\eqdef I_A-\mE^*(I_B)\ge0.
\ee
\bmyd
We say that a channel $\mM\in\cptp(A\to B)$ has the
\emph{reference-compatible completion property} if, for every
$\mE\in\cp_\le(A\to B)$ satisfying $\mE\le\gamma\mM$ for some
$\gamma\ge0$, there is a completely positive map $\mR_{\mE}\in\cp(A\to B)$
such that
\be\label{eq:compatible-completion-property}
 \mR_{\mE}^*(I_B)=\Delta_{\mE},
 \qquad
 \mR_{\mE}\le\|\Delta_{\mE}\|_\infty\mM\,.
\ee
\emyd
The first condition restores the missing trace; the second controls
the added map in CP order by the reference. Examples include:

\ben
\item When the input system is one-dimensional, channels are states.
For a subnormalized state $\widetilde\rho\le\gamma\sigma$, the trace
deficit is the scalar
$
\Delta=1-\tr\widetilde\rho
$,
and the missing mass can be restored by adding $\Delta\sigma$.
\item The property holds for classical channels when all smoothing
maps are restricted to the classical theory. Indeed, if
$M:X\to Y$ is a stochastic channel and $E:X\to Y$ is a classical
subchannel, define
\be
\Delta_x
\eqdef
1-\sum_yE(y|x)
\qquad\text{and}\qquad
R_E(y|x)
\eqdef
\Delta_xM(y|x).
\ee
Then $R_E$ restores the missing mass for each input and satisfies
\be
R_E\le
\left(\max_x\Delta_x\right)M\,.
\ee
Here it is important that the ambient theory is classical. This does
not imply that a classical channel, regarded as a quantum channel, has
the same property with respect to arbitrary quantum subchannels.
\item Every replacer reference $\mM(X)=\tr(X)\sigma_B$ has the
property in the full quantum theory. With $\Delta\eqdef\Delta_{\mE}$,
one may take
\be
\mR_{\mE}(X)
\eqdef
\tr(X\Delta)\sigma_B.
\ee
Indeed,
$
\mR_{\mE}^*(I_B)=\Delta
$
and
$
\mR_{\mE}
\le
\|\Delta\|_\infty\mM
$.
\een

More generally, the property holds for orthogonally flagged
block-replacer channels
\be
\mM(X)=\sum_a\tr(P_aX)\sigma_a,
\ee
where $\{P_a\}_a$ are mutually orthogonal projections summing to
$I_A$ and the states $\{\sigma_a\}_a$ have mutually orthogonal
supports. Indeed, $\mE\le\gamma\mM$ forces every Kraus operator of
$\mE$ to map each $P_aA$ into $\supp(\sigma_a)$, so
$\Delta_{\mE}$ is block diagonal with respect to $\{P_a\}_a$.
The map
\be
\mR_{\mE}(X)=\sum_a\tr(P_a\Delta_{\mE}P_aX)\sigma_a
\ee
then satisfies~\eqref{eq:compatible-completion-property}.

\bmyt
\label{thm:unified-completion-bound}
Let $\mN,\mM\in\cptp(A\to B)$, let $\delta\in(0,1)$, and
suppose that $\mM$ has the reference-compatible completion property.
Then
\be\label{eq:unified-completion-first}
D_{\max}^{\delta}(\mN\|\mM)
\le
\log \left(
2^{D_{\max}^{\delta,\le}(\mN\|\mM)}+\delta
\right).
\ee
Consequently, whenever $D_{\max}^{\delta,\le}(\mN\|\mM)$ is finite,
\be\label{eq:unified-completion-gap}
0
\le
D_{\max}^{\delta}(\mN\|\mM)
-
D_{\max}^{\delta,\le}(\mN\|\mM)
\le
\log \frac1{1-\delta}.
\ee
Moreover, the additive constant in~\eqref{eq:unified-completion-gap}
is optimal.
\emyt

\begin{proof}
Let $\mE\in\mb_\le^\delta(\mN)$ satisfy $\mE\le\gamma\mM$,
and write $\Delta\eqdef\Delta_{\mE}$. For a state $\rho_A$
with purification $\psi_{RA}$,
\be
0\le\tr(\rho_A\Delta)
=\tr\!\left[(\id_R\otimes(\mN-\mE))(\psi_{RA})\right]
\le\big\|(\id_R\otimes(\mN-\mE))(\psi_{RA})\big\|_+.
\ee
Taking the supremum gives
\be\label{eq:deficit-bound-unified}
\|\Delta\|_\infty\le\|\mN-\mE\|_{\diamond,+}\le\delta.
\ee
Let $\mR_{\mE}$ be a map satisfying
\eqref{eq:compatible-completion-property}, and define
$
 \mT\eqdef\mE+\mR_{\mE}
$.
The first condition in~\eqref{eq:compatible-completion-property}
makes $\mT$ trace preserving, while the second gives
\be\label{eq:unified-domination}
 \mT
 \le
 \bigl(\gamma+\|\Delta\|_\infty\bigr)\mM
 \le
 (\gamma+\delta)\mM.
\ee

It remains to check that completing the trace does not enlarge the
smoothing radius.  For arbitrary $\psi\in\pure(RA)$, set
\be
 X\eqdef(\id_R\otimes(\mN-\mE))(\psi),
 \qquad
 Y\eqdef(\id_R\otimes\mR_{\mE})(\psi)\ge0.
\ee
Because $\mR_{\mE}$ restores precisely the trace lost by $\mE$, we have
$\tr [Y]=\tr[ X]\ge0$.  It follows that
\ba
 \frac12\|X-Y\|_1
 &\le
 \frac12\|X\|_1+\frac12\|Y\|_1
 \\
 &=
 \frac12\|X\|_1+\frac12\tr [X]
 =\|X\|_+.
\ea
Since $\mN$ and $\mT$ are trace preserving, taking the supremum over
$\psi$ yields
\be
 \|\mN-\mT\|_{\diamond,+}
 \le
 \|\mN-\mE\|_{\diamond,+}
 \le\delta.
\ee
Thus $\mT\in\mb^\delta(\mN)$, and~\eqref{eq:unified-domination} gives
\be
 D_{\max}^{\delta}(\mN\|\mM)\le\log (\gamma+\delta).
\ee
Taking the infimum over all feasible $\mE$ and $\gamma$ proves
\eqref{eq:unified-completion-first}.

Equation~\eqref{eq:deficit-bound-unified} also implies
\be
 \mE^*(I_B)=I_A-\Delta\ge(1-\delta)I_A.
\ee
On the other hand,~$\mE\le\gamma\mM$ and the trace preservation
of $\mM$ imply $\mE^*(I_B)\le\gamma I_A$.  Hence
$\gamma\ge1-\delta$, and therefore
\be
 \gamma+\delta\le\frac{\gamma}{1-\delta}.
\ee
Together with the trivial inclusion
$\mb^\delta(\mN)\subseteq\mb_\le^\delta(\mN)$, this proves
\eqref{eq:unified-completion-gap}.

To see optimality, take $\mN=\mM$ and
$\mE=(1-\delta)\mN$.  Then
\be
 \|\mN-\mE\|_{\diamond,+}=\delta,
 \qquad
 D_{\max}(\mE\|\mN)=\log (1-\delta).
\ee
The lower bound $\gamma\ge1-\delta$ shows that this smoother is optimal. Thus
\be
 D_{\max}^{\delta,\le}(\mN\|\mN)=\log (1-\delta),
 \qquad
 D_{\max}^{\delta}(\mN\|\mN)=0,
\ee
and the upper bound in~\eqref{eq:unified-completion-gap} is attained.
\end{proof}

The bound in~\eqref{eq:unified-completion-gap} also holds for covariant channels. More precisely, let $G$ be a compact group, let $g\mapsto U_g$ be an irreducible
unitary representation on $A$, and let $g\mapsto V_g$ be a unitary
representation on $B$. Suppose that
\be
\mN(U_gXU_g^*)=V_g\mN(X)V_g^*
\qquad\text{and}\qquad
\mM(U_gXU_g^*)=V_g\mM(X)V_g^*\,,
\ee
for every $g\in G$. Then:
\bmyc
For every $\delta\in(0,1)$,
\be
D_{\max}^{\delta}(\mN\|\mM)
\le
\log \left(
2^{D_{\max}^{\delta,\le}(\mN\|\mM)}+\delta
\right),
\ee
and, whenever $D_{\max}^{\delta,\le}(\mN\|\mM)$ is finite,
\be
0\le
D_{\max}^{\delta}(\mN\|\mM)
-
D_{\max}^{\delta,\le}(\mN\|\mM)
\le
\log \frac1{1-\delta}\,.
\ee
\emyc

\begin{proof}
Let $\mE\in\mb_{\le}^{\delta}(\mN)$ satisfy
$\mE\le\gamma\mM$, and define its twirling by
\be
\overline{\mE}(X)
\eqdef
\int_G
V_g^*\mE(U_gXU_g^*)V_g\,dg,
\ee
where $dg$ is normalized Haar measure. Joint covariance of $\mN$ and $\mM$, together with convexity and unitary
invariance of the generalized diamond norm, implies both
$
\overline{\mE}\le\gamma\mM$ and
\be
\|\overline{\mE}-\mN\|_{\diamond,+}\le\delta\,.
\ee
The trace deficit
\be
\Delta\eqdef I_A-\overline{\mE}^{*}(I_B)
\ee
commutes with every $U_g$. Hence, by irreducibility and Schur's
lemma (see the symmetry methods in~\cite{Gour2025,Watrous2018}),
$\Delta=\alpha I_A$ for some $\alpha\ge0$. Moreover,
\be
\alpha=\|\Delta\|_\infty
\le\|\overline{\mE}-\mN\|_{\diamond,+}
\le\delta.
\ee
Thus $\mR_{\overline{\mE}}\eqdef\alpha\mM$ satisfies both
completion conditions for the twirled smoother. The proof of
Theorem~\ref{thm:unified-completion-bound} now applies: the channel
$\mT\eqdef\overline{\mE}+\alpha\mM$ remains within distance
$\delta$ of $\mN$ and obeys $\mT\le(\gamma+\delta)\mM$.
Taking the infimum and using $\gamma\ge1-\delta$ proves the claims.
\end{proof}

Whenever the hypotheses of Theorem~\ref{thm:unified-completion-bound}
or the corollary hold at every tensor power, the finite smoothed
quantities differ by at most $\log (1/(1-\delta))$. This bound
vanishes after division by the number of channel uses, so the two
smoothing conventions give the same asymptotic rates.

For general quantum channels, the trace deficit need not admit a
completion controlled by the reference. We demonstrate this obstruction
in two steps: a qubit example has infinite CPTP-smoothed max-relative
entropy but a finite subchannel-smoothed value; a qutrit family then
shows that the separation persists asymptotically even when the
max-relative entropy is finite.

\section{Quantum completion gaps: single-shot and asymptotic separations}
\label{sec:quantum-completion-gaps}

The preceding section identified several settings in which a subchannel
smoother can be completed to a trace-preserving channel at a uniformly
bounded cost. We now show that no analogous completion principle holds for
general quantum channels. We establish two complementary separations. First,
we construct a pair of qubit channels exhibiting a maximal single-shot gap:
for every prescribed nontrivial smoothing radius, the CPTP-smoothed
max-relative entropy is infinite, while the subchannel-smoothed quantity is
finite. This demonstrates that the obstruction already appears in the
smallest nontrivial quantum dimension. Second, we construct a pair of qutrit
channels for which the gap persists under tensor powers, yielding a strict
separation between the corresponding asymptotic rates. The latter provides
our main counterexample and shows that an arbitrarily small trace deficit can
produce an extensive advantage that cannot be recovered by imposing exact
trace preservation.

\subsection{A maximal single-shot separation for qubit channels}
\label{subsec:maximal-qubit-separation}

\bmyt
\label{thm:maximal-max-smoothing-failure}
Let $\eps\in(0,1)$ and $\delta\in[0,1)$. Then there exist qubit
channels $\mN$ and $\mM$ such that
\be\label{ball}
D_{\max}^{\delta}(\mN\|\mM)=\infty
\qquad\text{while}\qquad
D_{\max}^{\eps,\L}(\mN\|\mM)<\infty\,.
\ee
\emyt

\begin{remark}
Consequently, no universal inequality of the form
\be\label{eq:impossible-max-comparison}
D_{\max}^{\delta}(\mN\|\mM)
\le
D_{\max}^{\eps,\L}(\mN\|\mM)
\ee
can hold with $\delta<1$.
\end{remark}

\begin{proof}
We take $\mN\eqdef\id_2$ to be the identity qubit channel, and $\mM$ to be the qubit channel:
\be
\mM(X)\eqdef K_0XK_0^*+K_1XK_1^*.
\ee
where
\be\label{kraus}
K_0
\eqdef
\begin{pmatrix}
\sin(\alpha)&0\\
0&\sin(\beta)
\end{pmatrix}
\qquad\text{and}\qquad
K_1
\eqdef
\begin{pmatrix}
0&\cos(\beta)\\
\cos(\alpha)&0
\end{pmatrix}\,,
\ee
and $0<\alpha<\beta<\pi/2$ are some parameters to be determined shortly. By definition, $K_0^*K_0+K_1^*K_1=I_2$
so the operators in~\eqref{kraus} define a
qubit channel. Moreover, it is simple to verify that
the four operators $\{K_i^*K_j\}_{i,j=0}^1$ are linearly independent, and therefore,
by Choi's extremality criterion~\cite{Choi1975}, $\mM$ is an extreme channel.
This two-Kraus qubit family belongs to the standard extremal-channel
parametrization~\cite{RSW2002}; its use for the smoothing separation
is the point of the present construction.

We next choose $\alpha$ and $\beta$ such that $D_{\max}^{\eps,\L}(\mN\|\mM)<\infty$. By definition, this is equivalent of finding $\alpha$ and $\beta$ such that
$E_\gamma(\mN\|\mM)<\eps$ for sufficiently large, but finite $\gamma$.
Let $\psi\in\pure(RA)$ be arbitrary pure state and define $|\phi\ra\in RA$ to be the unnormalized vector
\be
|\phi\rangle
\eqdef
(I_R\otimes K_0)|\psi\ra\,.
\ee
Observe that the relation 
$K_0^*K_0\ge\sin^2(\alpha)I_2$
yields $\la\phi|\phi\ra\ge \sin^2(\alpha)$.
Since
$(\id_R\otimes\mM)(\psi)\ge\phi$, 
monotonicity of the hockey-stick divergence in its second argument
gives
\be\label{ega}
E_\gamma\left(
\psi\middle\|
(\id_R\otimes\mM)(\psi)
\right)
\le
E_\gamma\left(
\psi\middle\|
\phi
\right).
\ee

Let $x\eqdef\gamma \la\phi|\phi\ra\ge\gamma\sin^2(\alpha)$, and choose $\gamma>1/\sin^2(\alpha)$ so that $x>1$.
For $x>1$, a direct calculation on the span of $|\psi\rangle$ and
$|\phi\rangle$ gives
\be
E_\gamma\left(
\psi\middle\|
\phi
\right)
=
\frac{
1-x+\sqrt{(x-1)^2+4xy}
}{2}\qquad\text{where}\qquad y
\eqdef
1-
\frac{|\langle\psi|\phi\rangle|^2}{\la\phi|\phi\ra}\,.
\ee
Using
\be
\sqrt{(x-1)^2+4xy}
\le
(x-1)+\frac{2xy}{x-1}\,,
\ee
we obtain
\be\label{eq:counterexample-rank-one-bound}
E_\gamma\left(
\psi\middle\|\phi
\right)
\le
\frac{xy}{x-1}\,.
\ee
We next upper bound this expression.
Since the function $x\mapsto\tfrac{x}{x-1}$
is decreasing on $(1,\infty)$, and since
$x\ge\gamma\sin^2(\alpha)$
we obtain
\be\label{glim}
\frac{x}{x-1}\le\frac{\gamma \sin^2(\alpha)}
{\gamma \sin^2(\alpha)-1}\xrightarrow{\gamma\to\infty}1\,.
\ee
It remains to upper bound $y$. Let
$
\rho\eqdef\tr_R[\psi_{RA}]
$
be the reduced density matrix of $\psi_{RA}$. Then
\be
\la\phi|\phi\ra=\tr[\rho K_0^2]
\qquad\text{and}\qquad
\la\psi|\phi\ra=\tr[\rho K_0].
\ee
Thus,
\ba\label{yrhs}
y
&\le
\max_{\rho\in\md(\C^2)}
\left\{
1-
\frac{\left(\tr[\rho K_0]\right)^2}
{\tr[\rho K_0^2]}
\right\}
\\
&=
\left(
\frac{\sin(\beta)-\sin(\alpha)}
{\sin(\beta)+\sin(\alpha)}
\right)^2.
\ea
Indeed, since $K_0$ is diagonal, both
$\tr[\rho K_0]$ and $\tr[\rho K_0^2]$ depend only on the diagonal
entries of $\rho$. Hence it suffices to take
$\rho=\diag(p,1-p)$ with $p\in[0,1]$. The maximization over $p\in[0,1]$ yields the right-hand side of~\eqref{yrhs}.

From the upper bound in~\eqref{yrhs} and the limit in~\eqref{glim} we get that for $\beta$ sufficiently close to $\alpha$, there exists a finite $\gamma>1$ such
that $E_\gamma(\mN\|\mM)<\eps$.
By the definition of the smooth-before-Lorenz-extension max-relative
entropy,
\be
D_{\max}^{\eps,\L}(\mN\|\mM)\le\log(\gamma)<\infty\,.
\ee
It remains to prove that the range of $\alpha$ and $\beta$ can be chosen in such a way that in addition we have that the CPTP-smoothed max-relative entropy is
infinite. 

Indeed, by choosing sufficiently small $\alpha>0$ such that $\cos^2(\alpha)>\delta$ we get that
\ba
\frac12\|\mN-\mM\|_\diamond
&\ge
\frac12
\left\|
|0\rangle\langle0|
-
\mM(|0\rangle\langle0|)
\right\|_1
\\
&=
\cos^2(\alpha)>\delta\,,
\label{cedis}
\ea
where the equality follows from
\be
\mM(|0\rangle\langle0|)
=
\sin^2(\alpha)|0\rangle\langle0|
+
\cos^2(\alpha)|1\rangle\langle1|\,.
\ee
Now, let $\mE\in\cptp(\mathbb C^2\to\mathbb C^2)$ be $\delta$-close (in diamond distance) to $\mN$. From~\eqref{cedis} it follows that $\mE\ne\mM$.
Suppose, toward a contradiction, that
\be
s\eqdef 2^{D_{\max}(\mE\|\mM)}<\infty\,.
\ee
Then $s\ge 1$ and $\mE\le s\mM$.
If $s=1$ then $\mE=\mM$ and we get a contradiction.
Thus $s>1$. Define
\be
\mF
\eqdef
\frac{s\mM-\mE}{s-1}.
\ee
Then $\mF$ is completely positive and trace preserving, and
\be
\mM
=
\frac1s\,\mE
+
\left(1-\frac1s\right)\mF\,.
\ee
Since $\mM$ is extreme, this convex decomposition is trivial, so
again we get the contradiction $\mE=\mM$.
Therefore every channel $\mE$ that is $\delta$-close (in diamond distance) to $\mN$
obeys $D_{\max}(\mE\|\mM)=\infty$.
Taking the infimum over this $\delta$-ball gives $D_{\max}^{\delta}(\mN\|\mM)=\infty$.
This completes the proof.
\end{proof}

\bmyc
Let $0<\eps<1$. Then there exist qubit channels $\mN$ and $\mM$ such that
\be\label{eq:qubit-subchannel-smoothing-ball}
D_{\max}^{\eps}(\mN\|\mM)=\infty
\qquad\text{while}\qquad
D_{\max}^{\eps,\le}(\mN\|\mM)<\infty\,.
\ee
\emyc

\begin{proof}
Set
$\delta\eqdef\left(\sqrt{1+\eps/2}-1\right)^2$.
Then $\delta\in(0,1)$ and
$
4\sqrt{\delta}+2\delta=\eps
$.
By Theorem~\ref{thm:maximal-max-smoothing-failure}, applied with
Lorenz-smoothing parameter $\delta$ and CPTP-smoothing parameter $\eps$,
there exist qubit channels $\mN$ and $\mM$ such that
\be
D_{\max}^{\eps}(\mN\|\mM)=\infty
\qquad\text{while}\qquad
D_{\max}^{\delta,\L}(\mN\|\mM)<\infty\,.
\ee
On the other hand, Theorem~\ref{thm:subchannel-vs-Lorenz-max}, together
with $4\sqrt{\delta}+2\delta=\eps$, gives
\be
D_{\max}^{\eps,\le}(\mN\|\mM)
\le
D_{\max}^{\delta,\L}(\mN\|\mM)
<
\infty\,.
\ee
This proves the claim.
\end{proof}

The channels in the proof above satisfy
$D_{\max}(\mN\|\mM)=\infty$.
One can get the same result even with channels satisfying $D_{\max}(\mN\|\mM)<\infty$. That is, the obstruction remains arbitrarily strong even when the original
channel pair is required to have finite max-relative entropy.

\bmyt[Unbounded separation under finite $D_{\max}$]
\label{thm:finite-Dmax-unbounded-separation}
For every $\eps\in(0,1)$ and every $\delta\in[0,1)$,
\be\label{eq:unbounded-finite-Dmax-gap}
\sup_{\substack{
\mN,\mM\in\cptp(\mathbb C^2\to\mathbb C^2)\\
D_{\max}(\mN\|\mM)<+\infty
}}
\Big\{
D_{\max}^{\delta}(\mN\|\mM)
-
D_{\max}^{\eps,\L}(\mN\|\mM)
\Big\}
=
+\infty.
\ee
\emyt

\begin{proof}
Choose $\mN=\id_2$, $\mM$, and a finite $\gamma>1$ as in the
proof of Theorem~\ref{thm:maximal-max-smoothing-failure}, so that
\be
\frac12\|\mN-\mM\|_\diamond>\delta
\ee
and $E_\gamma(\mN\|\mM)<\eps$.
For $t\in(0,1)$, define
\be
\mM_{t}
\eqdef
(1-t)\mM+t\mN.
\ee
Since $\mM_{t}\ge t\mN$, we have
\be\label{eq:regularized-finite-Dmax}
D_{\max}(\mN\|\mM_{t})\le\log\left(\frac1t\right)<\infty\,.
\ee

Set
\be
s
\eqdef
\frac{\gamma}{1+(\gamma-1) t}.
\ee
and observe that $s\in(1,\gamma)$. Moreover,
\be
\mN-s\mM_{t}
=
\frac{1-t}{1-t+\gamma t}
\bigl(
\mN-\gamma\mM
\bigr)\,.
\ee
Thus, by positive homogeneity of the hockey-stick divergence,
\ba
E_{s}(\mN\|\mM_{t})
&=
\frac{1-t}{1-t+\gamma t}
E_\gamma(\mN\|\mM)
\\
&\le E_\gamma(\mN\|\mM)<\eps\,.
\ea
Consequently,
\be\label{eq:regularized-Lorenz-max-upper}
D_{\max}^{\eps,\L}(\mN\|\mM_{t})
\le
\log (s)
\le
\log(\gamma)\,.
\ee
The key point is that this upper bound does not depend on $t$.
On the other hand,
\ba
\lim_{t\to 0^+}D_{\max}^{\delta}
(\mN\|\mM_t)=D_{\max}^{\delta}
(\mN\|\mM)=\infty\,.
\ea
To justify this limit, suppose instead that some sequence $t_k\downarrow0$
has bounded smoothing cost. Choose CPTP smoothers $\mT_k$ in the
closed $\delta$-ball and a common finite $c$ with $\mT_k\le c\mM_{t_k}$.
The finite-dimensional set of channels is compact in its Choi
representation~\cite{Watrous2018}. A subsequence therefore converges
to a CPTP map $\mT$ with $\tfrac12\|\mT-\mN\|_\diamond\le\delta$
and $\mT\le c\mM$, contradicting the extremality argument above.
This completes the proof.
\end{proof}

\subsection{The cost of trace preservation on R\'enyi bounds}
\label{subsec:renyi-smoothing-bounds}

Let $\tD_\alpha$ denote the sandwiched R\'enyi divergence
\be
\tD_\alpha(\rho\|\sigma)
=\frac1{\alpha-1}\log\tr\!\left[
\left(\sigma^{\frac{1-\alpha}{2\alpha}}
\rho\,\sigma^{\frac{1-\alpha}{2\alpha}}\right)^\alpha\right],
\qquad \alpha>1,
\ee
with value $+\infty$ unless $\supp(\rho)\subseteq\supp(\sigma)$,
and its channel extension~\eqref{eq:prelim-channel-extension}.
Let $D_\alpha^{\mathbb M}$ denote the measured R\'enyi divergence,
obtained by maximizing the classical R\'enyi divergence over POVMs,
and use the same channel extension for this quantity.
For $\alpha>1$ and $0<\eps<1$, define
\be\label{eq:renyi-optimal-correction}
g_\alpha(\eps)=
\begin{cases}
\log(1-\eps),&\eps\ge1/\alpha,\\[1mm]
\displaystyle\frac{\log(1/\eps)+(\alpha-1)\log(\alpha-1)
-\alpha\log\alpha}{\alpha-1},&\eps<1/\alpha.
\end{cases}
\ee
The optimal state bound~\cite{Gour2026b} for the
modified smooth max-relative entropy is
$\widetilde D_{\max}^{\eps}(\rho\|\sigma)
\le D_\alpha^{\mathbb M}(\rho\|\sigma)+g_\alpha(\eps)$.
Applying this bound to every reference-assisted channel output gives
\ba\label{eq:renyi-Lorenz-bound}
D_{\max}^{\eps,\L}(\mN\|\mM)
&\le\big[D_\alpha^{\mathbb M}(\mN\|\mM)+g_\alpha(\eps)\big]_+\\
&\le\big[\tD_\alpha(\mN\|\mM)+g_\alpha(\eps)\big]_+,
\ea
where $[x]_+=\max\{0,x\}$. The positive part accounts for the
restriction $\gamma\ge1$ in Lorenz smoothing; the second inequality
uses $D_\alpha^{\mathbb M}\le\tD_\alpha$.
Set
\be
\eps_\delta=\left(\sqrt{1+\delta/2}-1\right)^2,
\qquad 0<\delta<1,
\ee
so that $4\sqrt{\eps_\delta}+2\eps_\delta=\delta$.
Theorem~\ref{thm:subchannel-vs-Lorenz-max} then yields
\ba\label{eq:renyi-subchannel-bound}
D_{\max}^{\delta,\le}(\mN\|\mM)
&\le\big[D_\alpha^{\mathbb M}(\mN\|\mM)
+g_\alpha(\eps_\delta)\big]_+\\
&\le\big[\tD_\alpha(\mN\|\mM)
+g_\alpha(\eps_\delta)\big]_+.
\ea
Thus the subchannel bound follows directly from Lorenz smoothing
and the uniform filter. A CPTP analogue fails already for a
simple qubit family.

\bmyl\label{thm:renyi-CPTP-separation}
There exists a family of qubit channel pairs $\mN,\mM_t$ with
finite $D_{\max}(\mN\|\mM_t)$ such that
\be\label{eq:renyi-CPTP-separation}
D_{\max}^{1/4}(\mN\|\mM_t)\xrightarrow{t\downarrow0}+\infty,
\qquad
\tD_\alpha(\mN\|\mM_t)\xrightarrow{t\downarrow0}2
\quad\text{for every fixed }1<\alpha<\infty.
\ee
Hence, for any fixed finite $\alpha>1$, no finite additive
correction depending only on the smoothing radius and system
dimensions can bound $D_{\max}^{\delta}$ by $\tD_\alpha$
uniformly over channel pairs.
\emyl

\begin{proof}
We reuse the Kraus operators $K_0,K_1$ in~\eqref{kraus}, writing
their angles as $\alpha',\beta'$ to reserve $\alpha$ for the
R\'enyi order in~\eqref{eq:renyi-subchannel-bound}.
For $0<t\le1/4$, choose
\be
\eta_t=e^{-1/t},\qquad
\alpha'=\frac\pi6,\qquad
\beta'=\sin^{-1}\!\left(\frac{1+\eta_t}{2}\right).
\ee
Set $\mN=\id_2$, $\mM(X)=\sum_{i=0}^1K_iXK_i^*$, and
$\mM_t=(1-t)\mM+t\mN$, as in the preceding regularization.
Here $\beta'$, and hence $K_0,K_1,\mM$, depend on $t$.
Since $\mN\le t^{-1}\mM_t$, the max-relative entropy is finite.

Let $\mT\le\gamma\mM_t$ be CPTP with
$\tfrac12\|\mT-\id\|_\diamond\le1/4$.
Linear independence of $I,K_0,K_1$ gives Kraus operators
$L_j=x_jI+y_jK_0+z_jK_1$, with
$X\eqdef\sum_j|x_j|^2\le\gamma t$ by CP domination.
Put $Y=\sum_j|y_j|^2$, $Z=\sum_j|z_j|^2$, and
$c=\sin(\alpha')\sin(\beta')\le3/8$.
Eliminating the cross term from the two diagonal
trace-preservation constraints gives $X=1+cY-(1+c)Z$.
The output probability of $|1\ra$ on input $|0\ra$ is
$\cos^2(\alpha')Z=3Z/4$, so $Z\le1/3$. Thus
\be
\gamma t\ge X\ge1-\frac{1+3/8}{3}=\frac{13}{24},
\qquad
D_{\max}^{1/4}(\mN\|\mM_t)\ge\log\frac{13}{24t}.
\ee

Fix $\alpha>1$ and $s=(\alpha-1)/\alpha$.
For any pure reference-assisted input $|\psi\ra$, set
$\sigma=(\id\otimes\mM_t)(\psi)$,
$|g\ra=2(I\otimes K_0)|\psi\ra$, and
$|e\ra=|\psi\ra-|g\ra$.
Then $\|g\|\le1+\eta_t$, $\|e\|\le\eta_t$, and
\be
\sigma\ge t\psi+\frac{1-t}{4}|g\rl g|,
\qquad
\|\sigma^{-1/2}\psi\|\le t^{-1/2},
\qquad
\|\sigma^{-1/2}g\|\le\frac2{\sqrt{1-t}}.
\ee
All inverse powers act on the support. H\"older's inequality,
$\|\sigma^{-s/2}u\|\le
\|u\|^{1-s}\|\sigma^{-1/2}u\|^s$,
and the triangle inequality give
\be
\|\sigma^{-s/2}\psi\|
\le(1+\eta_t)^{1-s}\left(\frac2{\sqrt{1-t}}\right)^s
+\eta_t^{1-s}
\left(\frac1{\sqrt t}+\frac2{\sqrt{1-t}}\right)^s.
\ee
The right-hand side tends to $2^s$, uniformly over $\psi$.
Since $\tD_\alpha(\psi\|\sigma)
=\frac2s\log\|\sigma^{-s/2}\psi\|$, this proves
\be
\limsup_{t\downarrow0}\tD_\alpha(\mN\|\mM_t)\le2.
\ee
The matching lower bound follows by using the input $|0\ra$:
\be
\tD_\alpha(\mN\|\mM_t)
\ge-\log\left(t+\frac{1-t}{4}\right)\longrightarrow2.
\ee
\end{proof}

This one-shot separation concerns a family of channel pairs;
the fixed-pair asymptotic threshold remains a separate question.

\subsection{An asymptotic separation for qutrit channels}
\label{subsec:asymptotic-qutrit-separation}

We now show that exact trace preservation can change the asymptotic
smoothing rate. The example compares the identity channel with a
mixture of the identity and a small phase rotation. Trace preservation
forces a polynomial identity that makes every nearby channel
expensive in CP order. Subchannels can approximate the identity
exponentially well without satisfying this identity and achieve
a strictly smaller rate.

For a unitary $U$, write $\mU(X)=UXU^*$.
CPTP smoothing uses $\tfrac12\|\cdot\|_\diamond$.
The subchannel construction works with both this distance and
the generalized diamond distance.

\bmyt\label{thm:qutrit}
Set
\be
t_0\eqdef(\sqrt5-2)^2\approx0.056,
\qquad
\theta_0\eqdef2\sin^{-1}(1/8)\approx0.25.
\ee
For $t\in(0,t_0)$ and $\theta\in(0,\theta_0)$, define
\be
U=\operatorname{diag}(1,e^{i\theta},e^{-i\theta})\,,
\qquad
\mN=\id_3\,,
\qquad
\mM=t\,\id_3+(1-t)\mU\,.
\ee
For every fixed $0<\delta<\sin^2\theta$,
\begin{align}\label{m1}
\lim_{n\to\infty}\frac1nD_{\max}^{\delta}
\left(\mN^{\otimes n}\big\|\mM^{\otimes n}\right)
&=\log\frac1t\,,\\
\limsup_{n\to\infty}\frac1nD_{\max}^{\delta,\le}
\left(\mN^{\otimes n}\big\|\mM^{\otimes n}\right)
&\le\frac12\log\frac1t+\log\frac4{1-t}
<\log\frac1t\,.
\label{m2}
\end{align}
\emyt

\begin{proof}[Proof of~\eqref{m1}]
For $S\subseteq[n]$, let $U_S$ act as $U$ at sites in $S$
and as $I$ elsewhere, and denote its unitary channel by $\mU_S$.
Since $I$ and $U$ are linearly independent, so are their tensor
products $\{U_S\}_{S\subseteq[n]}$. Moreover,
\be\label{oto}
\mM^{\otimes n}
=\sum_{S\subseteq[n]}t^{n-|S|}(1-t)^{|S|}\mU_S.
\ee

Suppose $\mT\le\gamma\mM^{\otimes n}$ is CPTP and
$\tfrac12\|\mT-\mN^{\otimes n}\|_\diamond\le\delta$.
Choose Kraus representations $\{L_j\}_j$ of $\mT$ and
$\{R_k\}_k$ of $\gamma\mM^{\otimes n}-\mT$.
Their combined family represents $\gamma\mM^{\otimes n}$.
By~\eqref{oto}, another representation consists of the linearly
independent operators
\be
K_S=\sqrt{\gamma t^{n-|S|}(1-t)^{|S|}}\,U_S.
\ee
The combined family is therefore an isometric mixing of
$\{K_S\}_S$~\cite{Watrous2018}:
\be
L_j=\sum_Sv_{j,S}K_S,
\qquad
R_k=\sum_Sw_{k,S}K_S.
\ee
Each column of the mixing matrix has squared norm one, so
\be
\sum_j|v_{j,S}|^2+\sum_k|w_{k,S}|^2=1,
\qquad
\sum_j|v_{j,S}|^2\le1.
\ee
Defining
\be
a_{j,S}\eqdef
\sqrt{\gamma t^{n-|S|}(1-t)^{|S|}}\,v_{j,S},
\ee
we obtain $L_j=\sum_Sa_{j,S}U_S$ and, in particular,
\be\label{gg}
\sum_j|a_{j,\varnothing}|^2
=\gamma t^n\sum_j|v_{j,\varnothing}|^2
\le\gamma t^n.
\ee

It suffices to show that
$\sum_j|a_{j,\varnothing}|^2\ge\kappa^2$
for some $\kappa>0$ independent of $n$ and of the smoother.
Indeed, this would imply $\gamma\ge\kappa^2t^{-n}$ and hence
\be\label{proofmt}
D_{\max}(\mT\|\mM^{\otimes n})
\ge n\log(1/t)+2\log\kappa.
\ee

Write $z^n\eqdef(z_1,\ldots,z_n)$ and define
\be
f_j(z^n)\eqdef
\sum_{S\subseteq[n]}a_{j,S}\prod_{i\in S}z_i,
\qquad
|f(z^n)\rangle\eqdef\sum_jf_j(z^n)|j\rangle,
\ee
where $\{|j\rangle\}_j$ is an orthonormal basis indexed by the
Kraus operators. Each variable has degree at most one.
The diagonal restriction
\be
|F(z)\rangle\eqdef|f(z,\ldots,z)\rangle
\ee
has constant coefficient
\be
|F(0)\rangle=\sum_ja_{j,\varnothing}|j\rangle,
\qquad
\langle F(0)|F(0)\rangle=\sum_j|a_{j,\varnothing}|^2.
\ee
We will bound this coefficient using normalization on the unit
circle and closeness to the identity at an interior point.

First, $|f(z^n)\rangle$ is normalized whenever
$z^n\in\{1,e^{i\theta},e^{-i\theta}\}^n$.
To see this, choose a normalized tensor product $|\psi\ra$
of eigenvectors of $U$ with respective eigenvalues $z_1,\ldots,z_n$.
Then $L_j|\psi\ra=f_j(z^n)|\psi\ra$, so trace preservation gives
\be
\langle f(z^n)|f(z^n)\rangle
=\sum_j|f_j(z^n)|^2
=\langle\psi|\sum_jL_j^*L_j|\psi\ra
=1.
\ee

This equality extends to the entire unit torus.
Fix all variables except one, denoted by $z$, and write
\be
|f(z^n)\rangle=|a\rangle+z|b\rangle.
\ee
For $|z|=1$, the expression
$\|\,|a\rangle+z|b\rangle\,\|_2^2-1$ has the form
\be
A+Bz+\overline Bz^{-1}.
\ee
If it vanishes at $1,e^{i\theta},e^{-i\theta}$, multiplying by
$z$ gives a polynomial of degree at most two with three distinct
roots. It must vanish identically. Starting with the other
variables at eigenvalues and extending one variable at a time
therefore yields
\be
\langle f(z^n)|f(z^n)\rangle=1
\qquad\text{whenever}\quad|z_1|=\cdots=|z_n|=1.
\ee
In particular,
\be
\|\,|F(z)\rangle\,\|_2=1
\qquad (|z|=1).
\ee

Next, let $|v\rangle$ be the equal superposition of the
eigenvectors corresponding to $e^{i\theta}$ and $e^{-i\theta}$.
Then
\be
\langle v|U|v\rangle=\cos\theta=:c,
\qquad 0<c<1.
\ee
For $|w\rangle\eqdef|v\rangle^{\otimes n}$,
\be
\langle w|U_S|w\rangle=c^{|S|},
\qquad
\langle w|L_j|w\rangle=f_j(c,\ldots,c).
\ee
Thus the probability that the output passes the test projecting
onto the input is
\be
\langle w|\mT(|w\rangle\langle w|)|w\rangle
=\sum_j|\langle w|L_j|w\rangle|^2
=\langle F(c)|F(c)\rangle.
\ee
The variational characterization of trace distance~\cite{Watrous2018}
now gives
\be
\begin{aligned}
1-\langle F(c)|F(c)\rangle
&\le\frac12
\bigl\||w\rangle\langle w|
-\mT(|w\rangle\langle w|)\bigr\|_1\\
&\le\frac12\|\mT-\mN^{\otimes n}\|_\diamond
\le\delta.
\end{aligned}
\ee
Consequently,
\be
\|\,|F(c)\rangle\,\|_2\ge\sqrt{1-\delta}.
\ee

To relate this to the constant coefficient, write
\be
|F(z)\rangle=|F(0)\rangle+z|G(z)\rangle,
\ee
where $|G(z)\rangle$ is a vector-valued polynomial.
On the unit circle,
\be
\|\,|G(z)\rangle\,\|_2
=\|\,|F(z)\rangle-|F(0)\rangle\,\|_2
\le1+\|\,|F(0)\rangle\,\|_2.
\ee
Applying the maximum-modulus principle~\cite{Rudin1987}
to $\langle u|G(z)\rangle$ for every unit vector $|u\rangle$
extends this bound to the disk. Hence
\be
\begin{aligned}
\|\,|F(c)\rangle\,\|_2
&\le\|\,|F(0)\rangle\,\|_2+c\|\,|G(c)\rangle\,\|_2\\
&\le c+(1+c)\|\,|F(0)\rangle\,\|_2.
\end{aligned}
\ee
Combining the two bounds gives
\be
\|\,|F(0)\rangle\,\|_2
\ge\frac{\sqrt{1-\delta}-c}{1+c}
=:\kappa>0,
\ee
where positivity follows from $\delta<1-c^2=\sin^2\theta$.
Since $\kappa$ depends only on $\theta$ and $\delta$,
this proves~\eqref{proofmt} uniformly over admissible smoothers.

For the upper bound, the identity itself is admissible and
$\mN^{\otimes n}\le t^{-n}\mM^{\otimes n}$. Therefore
\be
n\log(1/t)+2\log\kappa
\le D_{\max}^{\delta}(\mN^{\otimes n}\|\mM^{\otimes n})
\le n\log(1/t).
\ee
Dividing by $n$ and taking the limit proves~\eqref{m1}.
\end{proof}

\begin{proof}[Proof of~\eqref{m2}]
The parameter assumptions give
\be\label{h}
h\eqdef\|I-U\|_\infty=2\sin(\theta/2)<\frac14,
\qquad
4\sqrt t<1-t.
\ee
We approximate the identity by truncating the expansion
\be
I^{\otimes n}
=\bigotimes_{i=1}^n[U+(I-U)]
=\sum_{S\subseteq[n]}V_S,
\ee
where
\be
V_S\eqdef\bigotimes_{i=1}^nW_i^{(S)},
\qquad
W_i^{(S)}\eqdef
\begin{cases}
I-U,&i\in S,\\
U,&i\notin S.
\end{cases}
\ee
Retain only terms with at most $n/2$ factors $I-U$:
\be
E_n\eqdef\sum_{\substack{S\subseteq[n]\\|S|\le n/2}}V_S.
\ee
Since $\|V_S\|_\infty=h^{|S|}$,
\be
\begin{aligned}
\|E_n-I^{\otimes n}\|_\infty
&\le\sum_{k>n/2}\binom nk h^k\\
&\le h^{n/2}\sum_{k=0}^n\binom nk
=(2\sqrt h)^n
=:\eps_n.
\end{aligned}
\ee
The condition $h<1/4$ ensures that $\eps_n\to0$ exponentially.

Set
\be
N_n\eqdef\frac{E_n}{1+\eps_n},
\qquad
\mN_n(X)\eqdef N_nXN_n^*.
\ee
Then $\|N_n\|_\infty\le1$, so $\mN_n$ is a subchannel, and
\be
\|N_n-I^{\otimes n}\|_\infty
\le\frac{\|E_n-I^{\otimes n}\|_\infty+\eps_n}{1+\eps_n}
\le2\eps_n.
\ee
For any $X$,
\be
N_nXN_n^*-X=(N_n-I)XN_n^*+X(N_n^*-I).
\ee
This identity also holds after tensoring $N_n$ with the identity
on an arbitrary reference system. Using
$\|AXB\|_1\le\|A\|_\infty\|X\|_1\|B\|_\infty$
\cite{Bhatia1997} and taking the supremum gives
\be
\|\mN_n-\mN^{\otimes n}\|_\diamond
\le(1+\|N_n\|_\infty)\|N_n-I^{\otimes n}\|_\infty
\le4\eps_n.
\ee
In particular,
\be
\|\mN_n-\mN^{\otimes n}\|_{\diamond,+}
\le\|\mN_n-\mN^{\otimes n}\|_\diamond
\le4\eps_n\longrightarrow0.
\ee
Thus $\mN_n$ is an admissible subchannel smoother, whether
distance is measured by half the diamond norm or by the
generalized diamond norm, for every fixed radius $\delta>0$
and all sufficiently large $n$.

It remains to bound its CP-domination cost.
For an operator $K$, write $\Delta_K(X)=KXK^*$.
We use the elementary inequality
\be\label{cp-cs}
\Delta_{\sum_{j=1}^m\alpha_jK_j}
\le
\left(\sum_{j=1}^m|\alpha_j|^2\right)
\sum_{j=1}^m\Delta_{K_j}.
\ee
Indeed, the Choi operator of $\Delta_K$ is
$|K\rangle\!\rangle\langle\!\langle K|$~\cite{Choi1975},
and scalar Cauchy--Schwarz~\cite{Bhatia1997} gives
\be
\left|\sum_j\alpha_j\langle v|K_j\rangle\!\rangle\right|^2
\le
\left(\sum_j|\alpha_j|^2\right)
\sum_j|\langle v|K_j\rangle\!\rangle|^2
\ee
for every $|v\rangle$. This proves~\eqref{cp-cs} in CP order.

The definition of $\mM$ immediately gives
\be\label{bound-F}
\Delta_U\le\frac1{1-t}\mM.
\ee
For the other factor, write
\be
I-U
=\frac1{\sqrt t}(\sqrt t\,I)
-\frac1{\sqrt{1-t}}(\sqrt{1-t}\,U).
\ee
Applying~\eqref{cp-cs} to the two Kraus operators of $\mM$ yields
\be\label{bound-G}
\Delta_{I-U}
\le\left(\frac1t+\frac1{1-t}\right)\mM
=\frac1{t(1-t)}\mM.
\ee

Let $\mV_S(X)\eqdef V_SXV_S^*$.
Tensoring~\eqref{bound-F} and~\eqref{bound-G}, which preserves
CP order~\cite{Watrous2018}, gives
\be
\mV_S\le t^{-|S|}(1-t)^{-n}\mM^{\otimes n}.
\ee
For every retained term, $|S|\le n/2$, so
\be\label{bound-VS}
\mV_S\le t^{-n/2}(1-t)^{-n}\mM^{\otimes n}.
\ee

Write
\be
\mathcal S_n\eqdef\{S\subseteq[n]:|S|\le n/2\},
\qquad
m_n\eqdef|\mathcal S_n|\le2^n,
\qquad
\mE_n(X)\eqdef E_nXE_n^*.
\ee
Applying~\eqref{cp-cs} to $E_n=\sum_{S\in\mathcal S_n}V_S$
and then using~\eqref{bound-VS}, we obtain
\be
\begin{aligned}
\mE_n
&\le m_n\sum_{S\in\mathcal S_n}\mV_S\\
&\le m_n^2t^{-n/2}(1-t)^{-n}\mM^{\otimes n}\\
&\le4^nt^{-n/2}(1-t)^{-n}\mM^{\otimes n}.
\end{aligned}
\ee
Since $\mN_n=(1+\eps_n)^{-2}\mE_n\le\mE_n$,
\be
D_{\max}(\mN_n\|\mM^{\otimes n})
\le n\left(\frac12\log\frac1t+\log\frac4{1-t}\right).
\ee
Admissibility for all sufficiently large $n$ proves the first
inequality in~\eqref{m2}. The second follows from
$4\sqrt t<1-t$.
\end{proof}

\section{Equivalent formulations of the channel AEP}
\label{sec:equivalent-AEP-formulations}

In this section, we prove that four conjectured asymptotic
properties of a channel pair are equivalent: the fixed-error
channel Stein limit, the subchannel-smoothed AEP, a sharp threshold
for the channel hockey-stick divergence, and an AEP for the channel
Lorenz divergence. We establish their equivalence, rather than
their validity for general channel pairs. The distinction between
subchannel and CPTP smoothing is essential: the counterexamples
in Section~\ref{sec:quantum-completion-gaps} show that imposing
exact trace preservation can change the asymptotic smoothing rate.

The connection between asymptotic testing, smoothing, and divergence
thresholds originates in the information-spectrum approach
\cite{NH2007,DR2009,DL2015}. For channels, the relation between
hypothesis testing and smoothing the output states was established
in~\cite[Theorems~19--20]{FGW2025}. Our uniform subchannel filter
connects these descriptions to smoothing over a single
trace-nonincreasing map that approximates the channel on every
input, including entangled inputs. Together with the integral
representation of the Lorenz divergence, this yields the four
equivalent formulations.

The common threshold is the regularized channel relative entropy.
Below this threshold, we prove that the hockey-stick divergence
converges to one by applying the state Stein lemma to repeated
blocks of channel uses. Convergence to zero above the threshold
remains conjectural; we prove that it is equivalent to each of
the other three formulations.

Throughout this section, fix $\mN,\mM\in\cptp(A\to B)$ with
$D_{\max}(\mN\|\mM)<\infty$, and set
\be
\mu\eqdef D^{\mA}(\mN\|\mM)
=D^{\rm reg}(\mN\|\mM).
\ee
We use  the generalized-diamond subchannel
ball $\mb_\le^\delta(\mN)$ defined in
Section~\ref{sec:channel-vs-subchannel-smoothing}.
Recall from~\eqref{eq:prelim-channel-Lorenz} that
\be\label{eq:aep-Lorenz-integral}
D^\L(\mN\|\mM)
=\int_1^\infty\d\gamma
\left(\frac1\gamma E_\gamma(\mN\|\mM)
+\frac1{\gamma^2}E_\gamma(\mM\|\mN)\right).
\ee
For $n\in\N$, $r\in\R$, and $\eps\in(0,1)$, write
\ba\label{eq:aep-sequences}
\ell_n&\eqdef\frac1nD^\L
\left(\mN^{\otimes n}\middle\|\mM^{\otimes n}\right),\\
f_n(r)&\eqdef E_{2^{nr}}
\left(\mN^{\otimes n}\middle\|\mM^{\otimes n}\right),\\
h_n(\eps)&\eqdef\frac1nD_{\rm H}^\eps
\left(\mN^{\otimes n}\middle\|\mM^{\otimes n}\right).
\ea

We first establish the known below-threshold behavior of the
channel hockey-stick divergence and formulate the conjectured
above-threshold behavior. We then prove that this threshold
property is equivalent to the other three conjectures.

\subsection{Hockey-stick threshold behavior}
\label{subsec:hockey-stick-threshold}

We first isolate the part of the threshold behavior that holds
unconditionally. The following lemma records the direct part of channel Stein's
lemma~\cite{WW2019,FGW2025} and its formulation in terms of the
hockey-stick divergence. Both statements follow by applying the
state Stein lemma~\cite{HP1991,ON2000} to repeated blocks of
channel uses. We include the argument for completeness.
\bmyl\label{lem:aep-below-threshold}
For every $0\le r<\mu$,
\be\label{eq:aep-below-threshold}
\lim_{n\to\infty}f_n(r)=1.
\ee
Moreover, for every $\eps\in(0,1)$,
\be\label{eq:aep-direct-bound}
\liminf_{n\to\infty}h_n(\eps)\ge\mu.
\ee
\emyl
\begin{proof}
Fix $0\le r<\mu$ and choose $0<\delta<\mu-r$. By the definition of
$D^{\rm reg}$, there exist $k\in\N$ and
$\psi_{RA^k}\in\pure(RA^k)$ such that, with
\be
\rho\eqdef(\id_R\otimes\mN^{\otimes k})(\psi_{RA^k}),
\qquad
\sigma\eqdef(\id_R\otimes\mM^{\otimes k})(\psi_{RA^k}),
\ee
we have $D(\rho\|\sigma)>k(r+\delta)$. The state Stein lemma~\cite{HP1991,ON2000} gives
effects $0\le\Lambda_m\le I_{(RB^k)^{\otimes m}}$ satisfying
\be
\tr[\Lambda_m\rho^{\otimes m}]\longrightarrow1,
\qquad
\tr[\Lambda_m\sigma^{\otimes m}]\le2^{-mk(r+\delta)}
\ee
for all sufficiently large $m$. For $n=mk+s$, $0\le s<k$, use this
input and test on the first $mk$ channel uses, and discard the outputs
of the remaining $s$ uses. Then
\ba
f_n(r)&\ge\tr[\Lambda_m\rho^{\otimes m}]
-2^{nr}\tr[\Lambda_m\sigma^{\otimes m}]\\
&\ge\tr[\Lambda_m\rho^{\otimes m}]-2^{sr-mk\delta}
\longrightarrow1\,.
\ea
Since $f_n(r)\le1$, this proves~\eqref{eq:aep-below-threshold}.
The same tests are admissible for any fixed $\eps\in(0,1)$ once $n$
is large enough, and give
$\liminf_n h_n(\eps)\ge r+\delta$. Letting $r\uparrow\mu$
proves~\eqref{eq:aep-direct-bound} when $\mu>0$. When $\mu=0$,
the latter bound follows from $D_{\rm H}^\eps\ge0$.
\end{proof}

The Lorenz divergence measures the area under this threshold curve,
up to a term that vanishes after division by $n$.
\bmyl\label{lem:aep-area}
For every $n\in\N$ there exists
$\delta_n\in[0,1/n]$ such that
\be\label{eq:aep-area}
\ell_n=\int_0^\infty f_n(r)\,\d r+\delta_n\,.
\ee
In particular, $\liminf_{n\to\infty}\ell_n\ge\mu$.
\emyl
\begin{proof}
Substitute $\gamma=2^{nr}$ in the first term
of~\eqref{eq:aep-Lorenz-integral}, and define
\be
\delta_n\eqdef\frac1{n}\int_1^\infty
\frac1{\gamma^2}E_\gamma\left(\mM^{\otimes n}\middle\|\mN^{\otimes n}\right)\d\gamma.
\ee
Since $0\le E_\gamma\le1$, the claimed bounds on $\delta_n$ follow.
For $0\le r<\mu$, monotonicity gives
$\ell_n\ge r f_n(r)$. Apply
Lemma~\ref{lem:aep-below-threshold} and let $r\uparrow\mu$.
For $\mu=0$, the conclusion follows from nonnegativity.
\end{proof}

Thus both the hypothesis-testing rate and the Lorenz rate are at least
$\mu$ asymptotically. The next theorem shows that equality in either
case is precisely the missing above-threshold assertion.

\subsection{Equivalence of the four AEP conjectures}
\label{subsec:four-AEP-conjectures}

Let $\mN,\mM\in\cptp(A\to B)$ and suppose
$D_{\max}(\mN\|\mM)<\infty$. 
\bmyt\label{4conjectures}
The following statements are equivalent:
\ben
\item For every fixed $\eps\in(0,1)$, the regularized channel
hypothesis-testing divergence exists and satisfies
\be
\lim_{n\to\infty}\frac1nD_{\rm H}^\eps
\left(\mN^{\otimes n}\middle\|\mM^{\otimes n}\right)
=D^{\mA}(\mN\|\mM)\,.
\ee
\item The regularized channel Lorenz divergence exists and satisfies
\be
\lim_{n\to\infty}\frac1nD^\L
\left(\mN^{\otimes n}\middle\|\mM^{\otimes n}\right)
=D^{\mA}(\mN\|\mM)\,.
\ee
\item For every fixed $\delta\in(0,1)$, the regularized
subchannel-smoothed max-relative entropy exists and satisfies
\be
\lim_{n\to\infty}\frac1nD_{\max}^{\delta,\le}
\left(\mN^{\otimes n}\middle\|\mM^{\otimes n}\right)
=D^{\mA}(\mN\|\mM)\,.
\ee
\item For every $r>D^{\mA}(\mN\|\mM)$,
\be
\lim_{n\to\infty}E_{2^{nr}}
\left(\mN^{\otimes n}\middle\|\mM^{\otimes n}\right)=0\,.
\ee
\een
\emyt

\begin{proof}
We prove that each of the first three statements is equivalent to
statement~(4).
\ben
\item[\textbf{(a)}] $(1)\Longrightarrow(4)$.
Suppose that (1) holds, but $f_n(t)\not\to0$ for some $t>\mu$.
Then there exist $\delta\in(0,1)$ and a subsequence $\{n_j\}_{j\in\N}$ such that
$f_{n_j}(t)\ge\delta$ for all $j\in\N$. By the definition of the channel
hockey-stick divergence, choose
\be
\psi_{R_jA^{n_j}}\in\pure(R_jA^{n_j})\,,
\qquad R_j\simeq A^{n_j}\,,
\qquad 0\le\Lambda_{n_j}\le I_{R_jB^{n_j}}\,,
\ee
and define
\ba
p_{n_j}&\eqdef
\tr\!\left[\Lambda_{n_j}
(\id_{R_j}\otimes\mN^{\otimes n_j})
(\psi_{R_jA^{n_j}})\right]\,,\\
q_{n_j}&\eqdef
\tr\!\left[\Lambda_{n_j}
(\id_{R_j}\otimes\mM^{\otimes n_j})
(\psi_{R_jA^{n_j}})\right]\,,
\ea
so that
$
p_{n_j}-2^{n_jt}q_{n_j}\ge\delta
$.
Since $0\le p_{n_j},q_{n_j}\le1$, it follows that
\be
p_{n_j}\ge\delta,
\qquad
q_{n_j}\le2^{-n_jt}\,.
\ee
Thus, for $\eps\eqdef 1-\delta$, these tests are feasible at type-I error $\eps$, and
\be
h_{n_j}(\eps)
\ge-\frac1{n_j}\log  (q_{n_j})
\ge t>\mu\,,
\ee
contradicting (1).

$(4)\Longrightarrow(1)$.
Suppose (4) holds, and fix $\eps\in(0,1)$ and $t>\mu$.
For any
\be
\psi_{RA^n}\in\pure(RA^n)\,,
\qquad R\simeq A^n,
\qquad 0\le\Lambda_n\le I_{RB^n}\,,
\ee
define
\ba
p_n&\eqdef
\tr\!\left[\Lambda_n
(\id_R\otimes\mN^{\otimes n})(\psi_{RA^n})\right]\,,\\
q_n&\eqdef
\tr\!\left[\Lambda_n
(\id_R\otimes\mM^{\otimes n})(\psi_{RA^n})\right]\,.
\ea
If $p_n\ge1-\eps$, the definition of $f_n(t)$ gives
\be\label{gil}
f_n(t)\ge p_n-2^{nt}q_n\ge 1-\eps-2^{nt}q_n
\, .
\ee
By (4), $f_n(t)\to0$, so for all sufficiently large $n$ we have $f_n(t)\le(1-\eps)/2$. Combining with~\eqref{gil} we get
\be
q_n\ge2^{-nt}\bigl(1-\eps-f_n(t)\bigr)
\ge2^{-nt}\frac{1-\eps}{2}\,.
\ee
This bound holds for every feasible input and test. Hence
\be
h_n(\eps)
\le t+\frac1n\log \frac2{1-\eps}\,.
\ee
Taking the limsup and then letting $t\downarrow\mu$ yields
\be
\limsup_{n\to\infty}h_n(\eps)\le\mu\,.
\ee
Together with~\eqref{eq:aep-direct-bound}, this proves (1).

\item[\textbf{(b)}] $(2)\Longleftrightarrow(4)$.
Assume (2), and first suppose $\mu>0$. For $0<r<\mu<t$, the area bound~\eqref{eq:aep-area}
and nonnegativity of $f_n$ give
\be
\ell_n
\ge\int_0^\infty f_n(s)\,\d s
\ge\int_0^r f_n(s)\,\d s
+\int_r^t f_n(s)\,\d s\,.
\ee
Since $f_n$ is nonincreasing,
\be
f_n(s)\ge f_n(r)\quad(0\le s\le r)
\qquad\text{and}\qquad
f_n(s)\ge f_n(t)\quad(r\le s\le t)\,.
\ee
Consequently,
\be
\ell_n\ge r f_n(r)+(t-r)f_n(t)\,.
\ee
Taking the limit superior on both sides, using $\ell_n\to\mu$
and $f_n(r)\to1$ from~\eqref{eq:aep-below-threshold}, gives
\be
\mu\ge r+(t-r)\limsup_{n\to\infty}f_n(t)\,.
\ee
Rearranging gives
\be
\limsup_{n\to\infty}f_n(t)\le\frac{\mu-r}{t-r}\,.
\ee
Letting $r\uparrow\mu$ proves (4). If $\mu=0$, use instead
$\ell_n\ge t f_n(t)$.

Conversely, assume (4) and let $C\eqdef D_{\max}(\mN\|\mM)<\infty$.
The CP-order bound $\mN\le2^C\mM$ implies
$\mN^{\otimes n}\le2^{nC}\mM^{\otimes n}$, and hence
$f_n(r)=0$ for $r\ge C$. Hence all $f_n$ vanish outside the common interval $[0,C]$.

On this interval, Lemma~\ref{lem:aep-below-threshold} gives
convergence to one below $\mu$, while (4) gives convergence
to zero above $\mu$. Consequently,
\be\label{eq:aep-sharp-threshold}
\lim_{n\to\infty}f_n(r)=
\begin{cases}
1&\text{if }\; 0\le r<\mu\\
0&\text{if }\;r>\mu
\end{cases}
\ee
No convergence at $r=\mu$ will be needed for the following integral argument.

Since $0\le f_n(r)\le\1_{[0,C]}(r)$ on $[0,\infty)$, dominated
convergence~\cite{Rudin1987} yields
\be
\lim_{n\to\infty}\int_0^\infty f_n(r)\,\d r=\mu\,.
\ee
Equation~\eqref{eq:aep-area} now proves (2).
\item[\textbf{(c)}]$(3)\Longleftrightarrow(4)$.
We first record the comparison in the direction from subchannel
smoothing to the hockey-stick divergence. If
$\mE\in\mb_\le^\delta(\mN)$ and $\mE\le\gamma\mM$, then for every
input state $\psi$ and effect $\Lambda$,
\be
\tr\!\left[\Lambda(\id\otimes(\mN-\gamma\mM))(\psi)\right]
\le\tr\!\left[\Lambda(\id\otimes(\mN-\mE))(\psi)\right]
\le\delta\,.
\ee
Indeed, the output of $\mN-\mE$ has nonnegative trace, so the maximum
of the last expression over effects is its generalized trace norm.
Optimizing over inputs and tests gives
\be\label{eq:aep-subchannel-hockey-stick}
E_\gamma(\mN\|\mM)\le\delta\,.
\ee
Assume (3), and fix $r>\mu$ and $\delta\in(0,1)$.
For all sufficiently large $n$, (3) gives
\be
D_{\max}^{\delta,\le}
(\mN^{\otimes n}\|\mM^{\otimes n})<nr\,.
\ee
By the definition of the infimum, there is a subchannel
$\mE_n\in\mb_\le^\delta(\mN^{\otimes n})$ such that
\be
D_{\max}(\mE_n\|\mM^{\otimes n})<nr\,.
\ee
Equivalently, $\mE_n\le2^{nr}\mM^{\otimes n}$.
Applying~\eqref{eq:aep-subchannel-hockey-stick} with
$\mN^{\otimes n}$, $\mM^{\otimes n}$, $\mE_n$, and $2^{nr}$
in place of $\mN$, $\mM$, $\mE$, and $\gamma$, respectively, yields
\be
0\le f_n(r)
\le\|\mN^{\otimes n}-\mE_n\|_{\diamond,+}
\le\delta\,.
\ee
Thus $\limsup_{n\to\infty}f_n(r)\le\delta$.
Since $\delta\in(0,1)$ is arbitrary, $f_n(r)\to0$, proving (4).

For the converse, assume (4). Let $\delta\in(0,1)$ and let $\eps>0$ be such that $\delta=4\sqrt{\eps}+2\eps$. Theorem~\ref{thm:subchannel-vs-Lorenz-max}
then gives the generalized-diamond bound
\be\label{eq:aep-filtering-bound}
D_{\max}^{\delta,\le}(\mN\|\mM)
\le D_{\max}^{\eps,\L}(\mN\|\mM)\,.
\ee
For any $r>\mu$, (4) implies $f_n(r)\le\eps$ for all sufficiently
large $n$.  Thus $\gamma=2^{nr}$ is feasible in the definition of
$D_{\max}^{\eps,\L}$, and~\eqref{eq:aep-filtering-bound} gives
\be
D_{\max}^{\delta,\le}
\left(\mN^{\otimes n}\middle\|\mM^{\otimes n}\right)
\le
D_{\max}^{\eps,\L}
\left(\mN^{\otimes n}\middle\|\mM^{\otimes n}\right)
\le nr\,.
\ee
Dividing by $n$ and taking the limit superior yields
\be
\limsup_{n\to\infty}\frac1nD_{\max}^{\delta,\le}
\left(\mN^{\otimes n}\middle\|\mM^{\otimes n}\right)
\le r\,.
\ee
Letting $r\downarrow\mu$ proves the required upper bound.

For the lower bound, first suppose $\mu>0$ and fix $0\le r<\mu$.
By Lemma~\ref{lem:aep-below-threshold}, $f_n(r)\to1$.
Since $\delta<1$, we therefore have $f_n(r)>\delta$ for all
sufficiently large $n$.
For such $n$, no subchannel
$\mE\in\mb_\le^\delta(\mN^{\otimes n})$ can satisfy
$\mE\le2^{nr}\mM^{\otimes n}$. Indeed,
\eqref{eq:aep-subchannel-hockey-stick}, applied to
$\mN^{\otimes n}$, $\mM^{\otimes n}$, and $\gamma=2^{nr}$,
would imply
\be
f_n(r)
\le\|\mN^{\otimes n}-\mE\|_{\diamond,+}
\le\delta\,,
\ee
a contradiction. Thereofore, taking the infimum over admissible smoothers gives
\be
D_{\max}^{\delta,\le}
\left(\mN^{\otimes n}\middle\|\mM^{\otimes n}\right)
\ge nr\,.
\ee
Hence
\be
\liminf_{n\to\infty}\frac1nD_{\max}^{\delta,\le}
\left(\mN^{\otimes n}\middle\|\mM^{\otimes n}\right)
\ge r\,.
\ee
Letting $r\uparrow\mu$ proves the lower bound when $\mu>0$.

When $\mu=0$, let
$\mE\in\mb_\le^\delta(\mN^{\otimes n})$.
For every $\rho\in\md(A^n)$, the generalized diamond distance
bounds the trace deficit:
\be
0\le1-\tr[\mE(\rho)]
\le\|\mN^{\otimes n}-\mE\|_{\diamond,+}
\le\delta\,.
\ee
Thus
\be
\mE^*(I_{B^n})\ge(1-\delta)I_{A^n}\,.
\ee
If $\mE\le\gamma\mM^{\otimes n}$, trace preservation of
$\mM^{\otimes n}$ also gives
\be
\mE^*(I_{B^n})\le\gamma I_{A^n}\,,
\ee
so $\gamma\ge1-\delta$. Taking the infimum over $\gamma$
and admissible smoothers yields
\be
D_{\max}^{\delta,\le}
\left(\mN^{\otimes n}\middle\|\mM^{\otimes n}\right)
\ge\log (1-\delta)\,.
\ee
After division by $n$, the right-hand side tends to zero,
proving the lower bound also when $\mu=0$.
Together with the upper bound, this establishes (3).
\een
\end{proof}

Theorem~\ref{4conjectures} identifies the subchannel-smoothed AEP
with the strong converse for parallel channel Stein's lemma;
related output-divergence formulations appear in~\cite{FGW2025}.
The equivalence requires all fixed errors $\eps,\delta\in(0,1)$,
but no condition on $f_n(\mu)$. The finite-$D_{\max}$ assumption
provides the cutoff for the Lorenz integral; no support assumption
on the reversed pair is needed.
The same threshold property is also equivalent to
\be
\lim_{n\to\infty}\frac1nD_{\max}^{\eps,\L}
\left(\mN^{\otimes n}\middle\|\mM^{\otimes n}\right)=\mu
\qquad\forall\,\eps\in(0,1)\,.
\ee
Indeed, the level-$\eps$ crossing of the nonincreasing function
$f_n$ converges to $\mu$.

The CPTP-smoothed AEP would imply these equivalent statements,
by inclusion of the smoothing balls and the unconditional
subchannel lower bound. The reverse implication would require
control of trace-preserving completion. Accordingly, the CPTP
counterexample in Section~\ref{sec:quantum-completion-gaps}
does not settle the four conjectures for that pair: its subchannel
construction gives only an upper bound on the rate, not equality
with $\mu$.

\section{Failure of the CPTP-smoothed channel AEP}
\label{sec:failure-CPTP-smoothed-AEP}

For the qutrit pair $\mN,\mM$ of Theorem~\ref{thm:qutrit},
the CPTP-smoothed rate is $\log (1/t)$, whereas the subchannel
construction achieves a rate at most
\be\label{eq:CPTP-failure-rate}
R_t\eqdef\frac12\log \frac1t+\log \frac4{1-t}
<\log \frac1t\,.
\ee
We now show that $R_t$ also bounds the regularized channel
relative entropy from above. This disproves the CPTP-smoothed
AEP conjecture without assuming any of the equivalent conjectures in
Theorem~\ref{4conjectures}.

\bmyt[Failure of the CPTP-smoothed channel AEP]
\label{thm:CPTP-AEP-failure}
For the channel pair of Theorem~\ref{thm:qutrit},
\be\label{eq:CPTP-AEP-failure}
D^{\rm reg}(\mN\|\mM)\le R_t.
\ee
Consequently, the CPTP-smoothed AEP fails for this pair.
\emyt

\begin{proof}
The construction in the proof of Theorem~\ref{thm:qutrit}
provides subchannels $\mN_n$ satisfying
\be
\mN_n\le2^{nR_t}\mM^{\otimes n}\qquad\text{and}\qquad
\qquad
\|\mN^{\otimes n}-\mN_n\|_{\diamond,+}\xrightarrow{n\to\infty}0\,.
\ee
Applying~\eqref{eq:aep-subchannel-hockey-stick} to
$\mN^{\otimes n}$, $\mM^{\otimes n}$, and $\mN_n$,
with $\gamma=2^{nR_t}$, gives
\be\label{eq:CPTP-failure-hockey-stick}
0\le E_{2^{nR_t}}
\left(\mN^{\otimes n}\middle\|\mM^{\otimes n}\right)
\le\|\mN^{\otimes n}-\mN_n\|_{\diamond,+}
\xrightarrow{n\to\infty}0\,.
\ee
If $R_t<D^{\rm reg}(\mN\|\mM)$,
Lemma~\ref{lem:aep-below-threshold} would force this quantity
to converge to one. Thus $D^{\rm reg}(\mN\|\mM)\le R_t$,
strictly below the CPTP-smoothed rate in~\eqref{m1}.
\end{proof}

\section{Discussion and open problems}
\label{sec:discussion}
\label{subsec:AEP-counterexample-consequences}

In this paper, we showed that requiring a smoothing map to be
exactly trace preserving can change the leading asymptotic rate
and invalidate the channel AEP. The distinction is substantial:
for the qutrit family of Theorem~\ref{thm:qutrit},
\ba\label{eq:asymptotic-completion-gap-lower}
\liminf_{n\to\infty}\frac1n\Big[
&D_{\max}^{\delta}
(\mN^{\otimes n}\|\mM^{\otimes n})
-D_{\max}^{\delta,\le}
(\mN^{\otimes n}\|\mM^{\otimes n})
\Big]
\ge\log \frac{1-t}{4\sqrt t}\,,
\ea
for $0<\delta<\sin^2\theta$. This lower bound diverges as
$t\downarrow0$, with $\theta$ and $\delta$ fixed, although
each pair has finite $D_{\max}$. Meanwhile, the subchannel
approximation error vanishes exponentially even in generalized
diamond distance. An arbitrarily small trace deficit can therefore
have a lasting effect on the asymptotic rate.

Reducing the smoothing error does not restore the CPTP-smoothed
AEP. Indeed,
\be\label{eq:CPTP-small-error-failure}
\lim_{\delta\downarrow0}\lim_{n\to\infty}\frac1n
D_{\max}^{\delta}
(\mN^{\otimes n}\|\mM^{\otimes n})
=\log \frac1t>D^{\mA}(\mN\|\mM)\,.
\ee
The same rate holds along every sequence of nonnegative errors
$\delta_n\to0$. To see this, fix $0<\delta<\sin^2\theta$.
Eventually $\delta_n\le\delta$, so
\be
D_{\max}^{\delta}
(\mN^{\otimes n}\|\mM^{\otimes n})
\le D_{\max}^{\delta_n}
(\mN^{\otimes n}\|\mM^{\otimes n})
\le n\log (1/t)\,.
\ee
After dividing by $n$, the lower bound converges to
$\log (1/t)$ by~\eqref{m1}, while the upper bound equals
$\log (1/t)$. Hence
\be
\lim_{n\to\infty}\frac1nD_{\max}^{\delta_n}
\left(\mN^{\otimes n}\middle\|\mM^{\otimes n}\right)
=\log \frac1t\,.
\ee
Thus the CPTP-smoothed rate is unchanged, regardless of how
quickly the allowed error tends to zero.

This persistence of the gap reflects the cost of completing a
subchannel to a channel. Restoring the missing trace is always
possible, but it can substantially increase the max-relative
entropy relative to $\mM$. In the settings of
Section~\ref{subsec:subnormalization-harmless}, this increase
is bounded independently of the number of channel uses.
For the qutrit example, however, completing the constructed
subchannels while retaining a fixed small approximation error
requires an increase that grows at least linearly with the
number of channel uses.

The main question left open is whether the subchannel-smoothed
AEP holds for general channel pairs with finite $D_{\max}$.
By Theorem~\ref{4conjectures}, this is equivalent to
\be
E_{2^{nr}}(\mN^{\otimes n}\|\mM^{\otimes n})\longrightarrow0
\qquad
\forall\,r>D^{\mA}(\mN\|\mM)\,,
\ee
and to the strong converse for parallel channel Stein's lemma.
Our counterexample does not settle this question: even for the
qutrit pair, the construction bounds the subchannel rate by
$R_t$ without identifying it with $D^{\mA}(\mN\|\mM)$.

A complementary problem is to characterize when trace-preserving
completion has only a sublinear cost, with suitable control of
the smoothing error. Such a characterization would extend the
sufficient conditions established here. It would also be useful
to determine whether the asymptotic separation occurs for qubit
channels or for a fixed reference with positive-definite Choi
matrix. For the present qutrit family, determining the exact
subchannel rate and the CPTP-smoothed rate at larger fixed errors
would give a fuller account of how normalization affects smoothing.

\section*{Acknowledgments}
The author thanks the Guangdong Technion--Israel Institute of
Technology (GTIIT) for its warm hospitality and stimulating
research environment during the preparation of this work.
This research was supported by the Israel Science Foundation
under Grant No.~1192/24.

The author used GPT-6 Astra via Codex for assistance with
manuscript preparation and the development of the qutrit
counterexample. The author verified all arguments and takes
responsibility for the final manuscript.

\bibliographystyle{apsrev4-2}
\bibliography{AEP-references}

\end{document}